\documentstyle[preprint,aps,psfig]{revtex}
\newcommand{\be}{\begin{equation}}
\newcommand{\ee}{\end{equation}}
\newcommand{\ba}{\begin{eqnarray}}
\newcommand{\ea}{\end{eqnarray}}

\begin{document} 
\draft 
\title{ HAMILTONIAN SOLUTION OF THE  SCHWINGER MODEL WITH COMPACT U(1)} 
 
\author{Rom\'an Linares, Luis F. Urrutia and J. David 
Vergara \\ 
Departamento de F\'\i sica de Altas Energ\'\i as\\ 
Instituto de 
Ciencias Nucleares, \\ Universidad Nacional 
Aut\'onoma de M\'exico\\ 
Apartado Postal  70-543, 04510,  M\'exico D.F.}

\maketitle 
\begin{abstract} 
\baselineskip=12pt 
We present the complete exact solution of the  Schwinger model with compact gauge
group $U(1)$ (compact case). This is realized by demanding that the  true
electromagnetic  degree of freedom  $c$ has angular character. This is suggested by the loop approach to this
problem and  defines a  version of the Schwinger 
model which is different from the standard one, where the electromagnetic degree of freedom 
ultimately takes values on the line (non-compact case). All our results follow naturally from  the compactification condition. The main
consequences are: the spectra of the zero modes is not degenerated and does not correspond to the equally spaced
harmonic oscillator, the spectra and wave functions of the excited states also differ from those of the standard case,  both the electric charge and a modified gauge
invariant chiral charge are conserved and, finally, there is no need to
introduce a $\theta$-vacuum. Nevertheless, the axial-current anomaly
is still present. In more detail, these unusual properties turn out to be a consequence
of the following basic features: (i) the compactification condition
makes the   electromagnetic degree of freedom $c$ invariant under small and large gauge transformations. (ii) this full gauge invariance is  inherited by the fermionic
creation and  annihilation operators, which are subsequently used to solve the model
and (iii) the  boundary conditions upon the wave functions,which are imposed by demanding
hermiticity of both  the electric field and  the zero mode Hamiltonian, in the compact space of the variable $c$. They result in requiring
the wave functions $F(c)$ (and their first derivatives) to
be equal at points $\pm\, {\bar c}$, corresponding to the beginning and  the
end of the circle $-\, {\bar c} < \, c\, < \, {\bar c}$. These end points must be
identified  as a single physical point in the compact electromagnetic
configuration space. 
A comparison with the standard Schwinger model is  pointed out
along the text.
\end{abstract} 
\pacs{03.70, 11.15, 11.40.H} 
\maketitle 
\baselineskip=20pt

\section{Introduction}

One of the most popular  exactly soluble systems in quantum field theory 
is the Schwinger model, which describes electrodynamics in $1+1$ 
dimensions
\cite{Schwinger}. 
This model has been solved in many ways and we have not attempted here to provide a complete list of all the 
related references 
\cite{general,ADAMS,Capri,Manton,HetHo,Shifman,Link,IsoMurayama,HallinLiljenberg}.

Most of the solutions consider the one dimensional coordinate space as a line 
or a circle, with the basic  electromagnetic degree of freedom, the 
zero mode of the spatial component of the electromagnetic potential $A_1$, taking values in the interval $[ -\infty, + \infty]$. 
We refer to this version of the model as the non-compact case, 
where the topological  qualification refers, from now on, to the  electromagnetic 
configuration variables and not to the space coordinates. 

In general, gauge theories have also their compact version, which is 
realized when  the degrees of freedom 
corresponding to the gauge field  take values on a compact interval. 
The natural settings for this to happen are, on one hand,  the lattice 
formulation of gauge theories \cite{creutz}  and on the other, 
the loop-space formulation of gauge theories \cite{G-P}. In the latter 
approach, all the information about the theory is encoded in terms of
 variables which are  invariant under small and large gauge 
transformations. In $1+1$ dimensions the basic variable for the
electromagnetic degrees of freedom is $ \exp [ie \int_0^L dx A_1(x)]$ and, 
consequently, the loop representation naturally describes compact
electrodynamics, since it is enough to restrict $c= {1\over L} \int_0^L dx 
A_1(x)$ to the interval $\{-{\pi \over eL},{\pi \over eL}\}$ \cite{GMVU}. 
The loop representation 
can also encompass the 
non-compact case, at  the expense of introducing additional 
degrees of 
freedom \cite{FG1}. 

As emphasized by Polyakov \cite{Po}, the selection 
of one 
type of theory (compact case) over the other (non-compact case) has 
to be 
decided only on empirical grounds according to the predictions 
of each choice. From a more technical point of view we can say that 
compact gauge theories are blind to large gauge transformations, 
while non-compact ones have to be supplemented by $\theta$-vacua 
in order to preserve the invariance under such transformations. 
One  also expects that one of the main differences between the  compact 
and the non-compact cases 
would show up in the boundary conditions satisfied by the 
corresponding wave functionals, rather than in the specific 
form of the (functional) differential equations describing the dynamics . This is in complete 
analogy with the simple case of a one-dimensional particle in a 
line ( non-compact situation) versus the one dimensional rigid rotator 
(compact situation)\cite{Asorey}, which are both described by the same 
Schroedinger equation, but subjected to different boundary conditions. 


Many solutions to the standard Schwinger model, $c$ in the line $\{-\infty, +\infty\}$, start from
considering $c$ as an angular variable. Nevertheless, using appropriate boundary conditions, the corresponding authors manage to unfold the circle into the line, i.e. to go from compact $U(1)$ to its
universal covering \cite{Manton,Shifman,Link,IsoMurayama}. 

In this work we maintain the angular character of $c$ and fully explore the consequences of this choice which, up to our knowledge, are not previously reported in the literature. It is important to emphasize that our results follow uniquely from the compactification
condition, together with the standard definitions of both a scalar product and the hermiticity conditions in the corresponding Hilbert space. 
Not surprisingly, the compactification prescription leads to a model which drastically differs from the NCSM, as will be seen along the text. 

The compactification  of the gauge group $U(1)$ is realized by demanding that the only surviving electromagnetic degree
of freedom $c$ behaves as an angular variable living in a circle of
length $\frac{2\,\pi}{e L}$.  
In the sequel we call this model  the  compact
Schwinger model, or the compact case.  
Furthermore, we are going to take full advantage of one of the many 
methods which have been successfully applied previously to 
solve the non-compact  model. To this purpose we have selected the 
Hamiltonian approach of Refs. \cite{Link,IsoMurayama} upon which the 
present work heavily relies. 

In the present Hamiltonian formulation we obtain the complete solution 
of the  compact Schwinger model, including the ground 
and excited states.  A partial solution of the CSM was found in Ref. \cite{GMVU},  using the loop approach to this problem \cite{G-P}, and served as motivation for the work presented here. These partial results coincide with those obtained in this work. Previous progress in the solution of this model were
reported in \cite{PREREP}. 

The possibility of  rewriting the
initially  linear fermionic Hamiltonian in terms of the corresponding Sugawara currents \cite{SUGA}, together with the introduction of 
the Bogoliubov transformation in the Hamiltonian approach  are
the fundamental keys to obtain the full solution of the model.
The formulation of these  topics in the loop approach provides 
an interesting subject for future studies. 
 
The paper is organized as follows: Section II contains a brief motivation for our compactification procedure, arising from the loop approach  quantization of  the free electromagnetic field.
In Section III we define the  compact Schwinger model and state our notation and conventions. 
In Section IV  we discuss the gauge invariance of the  model. There it is shown how the
compactification  condition, i.e. our choice for the topology of $c$,  leads to fully gauge invariant electromagnetic and fermionic degrees of freedom, which are subsequently used in the resolution of the model. We also
consider the Gauss law and show that it implies 
that the physical states of the theory must be independent of the excited modes of the electromagnetic potential.
In section V we construct the fermionic Fock space in a background
electromagnetic field and  introduce the total electric charge $Q$ together with the total 
modified chiral charge $\bar Q_5$. Both charges are gauge invariant and conserved, in contrast with the non-compact case where $\bar Q_5$ is not fully gauge invariant. 
 In Section VI we perform 
the full quantization of the model. Using a Bogoliubov transformation 
we get a complete solution for the ground and excited states. In
Section VII we present a detailed proof of the conservation
of the modified chiral charge ${\bar Q}_5$. Accordingly, we show that there is  no spontaneous 
symmetry braking, since the fermionic condensates 
$\langle \bar \psi \psi \rangle$ and $\langle \bar \psi
\gamma_5 \psi \rangle$ are exactly zero. In consequence,  the $\theta$-vacuum structure is not required in the
compact Schwinger model. We summarize our results
in Section VIII, emphasizing the main differences with the standard Schwinger model. The Appendix contains the calculation of the regularized current algebra of the model, together
with the corresponding hermiticity properties. 

\section{Motivation}

In this section we provide a motivation for the compactification of the electromagnetic
degree of freedom in the framework of the loop-space formulation of the free electromagnetic field.
We take the space as a circle of length $L$ and impose periodic boundary conditions on $A_\mu$.

Classically, our choice of gauges is $A_0=0$, together with   $\partial_1 A_1=0$. This  implies $A_1=A_1(t)=c(t)$, leaving the zero mode of the electromagnetic potential as the only true electromagnetic degree of freedom,  with $E={\dot c}$.
We still have the freedom of large  gauge transformations $A_\mu \rightarrow A_\mu - \partial_\mu\, \alpha, \
\alpha= \frac{2\,\pi\,n}{e \,L}\, x $,
which change
\be
c(t)\rightarrow c(t) - \frac{2\,\pi\,n}{e \,L}\, 
\ee
leaving $A_0=0$ together with $E$ invariant.

The quantization is performed in the subspace $A_0=0=\Pi_0$, and subsequently imposing the remaining Gauss law constraint upon the physical states.
The Hamiltonian density,  the Gauss law  and the commutation relations  are
\ba
&&{\cal H}= \frac{1}{2}\, E^2, \qquad {\cal G}=\partial_x\, E \approx 0, \qquad  [A_1(x), \, E(y) ]=i\,\hbar\,  \delta(x-y),
\ea
Expanding in modes, we have
\ba
&&A_1(x,t)=c(t) +\sum_{m\neq0}A_m \,{\rm exp}\left( \frac{2\,\pi
\, m}{L}\,x \right),\nonumber \\
&&E(x,t)=E_0(t) +\sum_{m\neq0}E_m \, {\rm exp}\left( -\frac{2\,\pi
\, m}{L}\,x \right),\qquad E^2=\sum_{m \geq0}E_m^*\, E_m,
\ea
leading to
\ba
[A_m, \, E_n]=\frac{i\, \hbar}{L}\, \delta_{mn}
\quad \Rightarrow\qquad [{ c}(t),\,{ E}_0(t) ]=\frac{i \,\hbar }{L}.
\ea
In the connection representation we have the realization $E_n=\frac{i\, \hbar}{L} \frac{\partial}{\partial\, A_n}$ for the above commutators.
Only the zero-mode survives because the Gauss law implies $E_m=0, m\neq 0$, upon the physical states. That is to say $\Psi_{\rm phys}=\Psi(c)$.
The Hamiltonian is
\be
H= -\frac{\hbar^2}{2 L}\frac{d^2}{d\, c^2},
\ee
with the corresponding eigenfunctions  and eigenvalues
\be
\Psi(c)=\Psi_0\, {\rm exp}(i \,p\, c), \qquad E_p= \frac{\hbar^2\,p^2}{2\,L} 
\ee
The precise nature of the spectra will depend on the boundary conditions imposed upon the wave functions.

Now we turn to the loop approach, where we introduce
\be
T^0[\gamma]={\rm exp}\left( i\, e\, \oint_\gamma d\,x\, A_1(x,t) \right)=
{\rm exp}\left( i\, e\, c(t)\, M\, L \right),
\ee
as the manifestly gauge invariant
electromagnetic degree of freedom. Here $M=0, \,\pm\, 1, \, \pm\, 2 \dots$ is the number of turns of the loop $\gamma$, which completely characterize any loop in our spatial circle. In fact, $T^0[\gamma]$ is also invariant under large gauge transformations, because
\ba
T^0[\gamma]\rightarrow {\rm exp}\left( i\, e\, M\, L\, \left(c(t) - \frac{2\,\pi \, n}{e\,L}\right) \right)={\rm exp}\left( i\, M\, n\, 2\pi\right)T^0[\gamma]= T^0[\gamma].
\ea

Let us consider the  gauge invariant operators ${\hat T}^0[N]$ and ${\hat E}$, such that 
\be
{\hat T}^0[N]={\rm exp}\left( i\, e\, {\hat c} \, N\, L \right), \quad [{\hat T}^0[N],\, {\hat E}]=-\, e\,\hbar\, N\, {\hat T}^0[N].
\ee

The Hilbert space $\{ |N\rangle\}$, where the states are labeled by the number of turns, is built by
starting from  the vacuum $|0\rangle$ 
\ba
{\hat E}\, |0\rangle= 0, \qquad |N\rangle={\hat T}^0[N]\, |0\rangle, 
\ea
where we have
\ba
|N+P\rangle={\hat T}^0[N]\, |P\rangle , \quad {\hat E}\,|N \rangle=\, e\, N \,|N\rangle.
\ea
The Hamiltonian is
\be
H=\int_0^L d\,x \frac{1}{2} E^2, \qquad H\,|N\rangle=\frac{1}{2}(\,e^2\,\hbar^2\,L)\, N^2\, |N\rangle=(\,\hbar\, e^2\, L^2)\, \frac{\hbar}{2\, L}\, N^2\,|N\rangle.
\ee 

Here, the choice of boundary conditions, which determine the eigenvalues and eigenfunctions,  is somewhat hidden in the algebraic process. In order to make them explicit, we introduce a {modified} connection representation basis $|{\tilde c}\rangle$  defined by
\be
|{\tilde c} \rangle= \sum_N\, {\rm exp}(i\,e\,{\tilde c}\, N\,L)\,|N\rangle= 
\sum_N\, {\rm exp}\left(i\frac{\pi {\tilde c}}{{\bar c}} N\right)\,|N\rangle, \qquad  \quad {\bar c}= \frac{\pi}{e\, L},
\ee
which is a superposition of fully gauge invariant states. From the above expression  we verify that $|{\tilde c} + 2\, M\, {\bar c} \rangle\, =\, |{\tilde c} \rangle$, which states the angular character of the variable  ${\tilde c} $ in an interval of length $2\, {\bar c}$, which we choose
to be $-{\bar c} \leq
{\tilde c}\leq {\bar c}$. 

The inversion of the above relation is
\be
|N\rangle =\int_{-{\bar c}}^{{\bar c}} d\, {\tilde c}\, \,  
{\rm exp}\left(-i\frac{\pi {\tilde c}}{{\bar c}}\, N \right)\, |{\tilde c}\rangle.
\ee

Summarizing, in the modified connection representation with ${\tilde c}$ being the angular variable defined above,  we have the following realization for the operators 
\be
{\hat E}= \frac{\hbar}{i\, L}\frac{\partial}{\partial
{\tilde c}}, \qquad H=-\frac{\hbar^2}{2 L}\frac{\partial^2}{\partial
{\tilde c}^2},
\ee
together with the corresponding eigenfunctions, and its derivatives
\be
\Psi_N({\tilde c})=\frac{1}{\sqrt{2\,{\bar c}}}\,{\rm exp}\left(-i\frac{\pi {\tilde c}}{{\bar c}}\, N \right),\qquad \frac{d\, \Psi_N({\tilde c})}{d\,{\tilde c}}=-\frac{i\,\pi\,N}{\sqrt{2\,{\bar c}^3}}\,{\rm exp}\left(-i\frac{\pi {\tilde c}}{{\bar c}}\, N \right).
\ee
Clearly the above functions satisfy the boundary conditions
\ba
\Psi_N|_{{\tilde c}=-{\bar c}}=\Psi_N|_{{\tilde c}=+{\bar c}}, \qquad 
\frac{d\, \Psi_N}{d\,{\tilde c}}|_{{\tilde c}=-{\bar c}}=\frac{d\, \Psi_N}{d\,{\tilde c}}|_{{\tilde c}=+{\bar c}}.
\ea 
From now on we will denote by $c$ the compacted variable ${\tilde c}$.

\section{The model} 
 
In the sequel we use  similar notation and conventions 
as those in 
Ref.\cite{IsoMurayama}. 
The model is described by the Lagrangian 
 
\begin{equation} 
{\cal L} = -\frac{1}{4} F_{\mu\nu}F^{\mu\nu} 
           +\bar{\psi} \gamma^{\mu} \left(i\partial_{\mu} 
           - e A_{\mu}\right)\psi 
          \label{LAG} 
\end{equation} 
where $F_{\mu\nu}=\partial_{\mu}A_{\nu} -\partial_{\nu}A_{\mu}$, 
 $\bar{\psi}=\psi^\dagger \gamma^0$ is a  Grassmann valued 
fermionic field and  we are using units such that $\hbar=c=1$. 
We consider the coordinate space  to be $S^1$ and  we will require 
periodic(antiperiodic) 
boundary conditions for the fields 
 
\begin{equation} 
A_{\mu}(x+L) = A_{\mu}(x), \quad 
\psi(x+L) = -\psi (x), 
\label{BC} 
\end{equation} 
where  $L$ is  the length of the circle. 
The gamma matrices are: 
$\gamma^0=\sigma_1,\; \gamma^1=i\sigma_2,\; \gamma^5=-\gamma^0\gamma^1= 
\sigma_3$, where $\sigma_i$ are the standard Pauli matrices. 
We use the  signature $(+,-),\;\; i.e.\;\; \eta_{00}=-\eta_{11}=1$. 

 
After the standard canonical analysis of the Lagrangian density 
(\ref{LAG}), describing the configuration space variables 
$A_0,\,  A_1$ and $\psi$, we obtain
 
\begin{equation} 
{\cal H} = \frac{1}{2} E^2 
           + i \psi^\dagger\sigma_3 
           \left(\partial_1 + i e A_1\right)\psi 
              - A_0\,\left(\partial_1\, E - e\, \psi^\dagger \,
 \psi \right), \qquad \Pi_0\approx 0, 
\label{HAMDEN} 
\end{equation} 
where the  corresponding  canonical momenta are
$\Pi_0,\, \Pi_1=F_{01}=E $ and $\Pi_\psi=-i\, \psi^*$. Conservation
in time of the primary constraint $\Pi_0\approx 0$
leads to the Gauss law constraint
 
\begin{equation}\label{GL} 
{\cal G}=\partial_1 E - e \psi^\dagger\psi\approx 0. 
\end{equation} 
There are no additional constraints. 

At this stage we partially fix the gauge in
the electromagnetic potential by choosing 
\be
\label{GCHOICE}
 A_0=0, \quad  \Pi_0=0.
\ee
The only remaining constraint ${\cal G}$ is first class and
it will be imposed strongly upon the physical states of the system.

From now on we use the notation $A_1=A$ for the surviving
electromagnetic degree of freedom. Also we have 
$\psi = (\psi_1,\psi_2)^\top$, where 
$\top$ denotes transposition. The charge density is 
given by $\rho(x)= e \psi^\dagger \psi= e (\psi_1^*\psi_1 + 
 \psi_2^*\psi_2$). The resulting Poisson brackets algebra at equal times  is 
 
\begin{eqnarray} 
\left\{ A(x), E(y) \right\} = \delta(x,y), \qquad  
\left\{ \psi_{\eta}(x), \psi^*_{\xi}(y) \right\} 
      = -i \delta_{\eta \xi} 
          \delta(x,y), \quad \eta, \xi=1,2\,, 
\label{PB} 
\end{eqnarray} 
with all other brackets being zero. 
 
The standard canonical quantization procedure leads to the 
following non-zero commutators 
  $ (\hbar=1)$ 
 
\begin{eqnarray} 
\left[ A(x), E(y) \right] = i \delta(x,y), \qquad  
\left\{ \psi_{\eta}(x), \psi^*_{\xi}(y) \right\} 
      =  \delta_{\eta \xi} 
          \delta(x,y). 
\label{CCR} 
\end{eqnarray} 


The  gauge chosen in (\ref{GCHOICE}) does not completely fix the electromagnetic degrees of freedom, leaving  the Lagrangian density (\ref{LAG}) still invariant under the 
following gauge transformations 
 
\begin{equation}  
  \psi \rightarrow  {\rm e}^{ie \alpha(x)} \psi, 
  \quad A_\mu \rightarrow A_\mu - \partial_\mu \alpha (x), \label{GGT} 
\end{equation} 
generated by the remaining Gauss law constraint. The constant piece $\alpha_0$ of the function $\alpha(x)$ is 
irrelevant in the above transformation. 
In the sequel we consider ${ \bar \alpha}(x)= \alpha(x) - \alpha_0$ 
as the function generating the gauge transformations. Notice that $\partial_0 \alpha (x)=0$ respects the gauge condition $A_0=0$. 
 
There are two families of gauge transformations:
\begin{enumerate}\item Those continuously 
connected to the identity, 
called small gauge transformations (SGT), characterized by  the  function 
\begin{equation}
{\bar \alpha}(x) = b \left(e^{i 2 \pi n x  / L} -1  \right),
\end{equation}
which is periodic in $x$ and 
preserves 
the boundary conditions 
(\ref{BC}). 
\item The second family corresponds to the so called large gauge 
transformations (LGT), which are generated  by the 
non-periodic functions
\begin{equation}
\label{LGT}
  {\bar \alpha}(x)={2 \pi 
n \over e L}x=2\,n {\bar c}\,x , \quad n=\pm 1, \pm 2, \dots .
\end{equation}
The boundary conditions (\ref{BC}) are also preserved in this case. 
\end{enumerate} 

Notice that in both cases we have
\be
\label{ALFCOND}
{\bar \alpha}(0)=0.
\ee
At this stage we {\bf define} the compact Schwinger model  by demanding that 
the only true degree of freedom arising from the electromagnetic 
potential in one dimension, which is the zero mode 
$c$, be restricted 
to the interval
\begin{equation}\label{COMPACT} 
- {\bar c}\leq\, c=\frac{1}{L}\int_0^L A(z)\, dz \, \leq {\bar c}. 
\end{equation} 
In the previous section we provided
some motivation
for considering this situation. 

Equation (\ref{COMPACT})  means that two values of $ c $ differing by
$2\,{\bar c}\, N= \frac{2\, \pi\, N}{e L}$ 
must be  identified as a single physical point in the compact space of the variable $c$.  

Since the remaining electromagnetic degrees of freedom are pure gauge,
we expect the possible compactification of them to be
irrelevant. Nevertheless, there still remains the question of  in
which way, if any, the 
imposed compactification shows itself in the fermionic degrees of
freedom of the problem. These 
topic will be addressed to in the next section.  
 
\section{Identification of the gauge invariant degrees of freedom} 
 
Let us consider the following Fourier decomposition for the 
electromagnetic potential $A(x)$, the field strength $E(x)$ 
and the gauge transformation function ${\bar \alpha (x)}$ 
\begin{eqnarray}\label{RFD} 
 A(x) &=& c + \sum_{m \neq 0 } A_m \ { \rm e}^{\frac{2 \pi  i m 
}{L} x}, 
\quad 
E(x) = E_0+ \sum_{m \neq 0} E_m \ { \rm e}^{- \frac{2 \pi  i m }{L} x}, 
\nonumber \\ 
 {\bar \alpha}(x) &=& \sum_{m \neq 0} 
{\bar \alpha}_m \ { \rm e}^{\frac{2 \pi  i m }{L} x}, 
\end{eqnarray} 
leading to the following inverse transformations 
\begin{equation}\label{IRFD} 
c=\frac{1}{L} \int_0^L dx \ A(x), \quad A_m=\frac{1}{L} 
\int_0^L dx \ 
A(x) \   { \rm e}^{- \frac{2 \pi  i m }{L} x}, \ m\neq 0, 
\end{equation} 
with similar expressions for $E_0, E_m, {\bar  \alpha}_m $. Let us 
observe that $E_m^\dagger= E_{-m}$, 
$A_m^\dagger= A_{-m}$ in virtue of the hermiticity of both $E(x)$ 
and $A(x)$. Let us remark that 
\begin{equation}\label{ALFA0} 
0={\bar \alpha}(0) =  \sum_{m \neq 0} {\bar \alpha}_m, 
\end{equation} 
which will be used in the sequel.

Under a gauge transformation $ A(x) \rightarrow A(x) - 
\frac{\partial {\bar  \alpha}(x)}{\partial x} $, 
the corresponding modes change as 
\begin{equation}\label{GTOM} 
c \rightarrow c -\frac{1}{L} ({\bar \alpha} (L) - {\bar \alpha}(0)) , 
\quad A_m \rightarrow A_m - 
 \frac{2 \pi i m}{L} \ {\bar \alpha}_m, \quad m\neq 0 . 
\end{equation}
 
Clearly, the zero mode $c$ is invariant under  the small gauge 
transformations 
generated by ${\bar \alpha}= b\left( {\rm e}^{i 2 \pi n x/L}-1 
\right)$. As for the LGT (\ref{LGT}), $c \rightarrow c - \frac{2 
\pi n}{e L 
}$, but these points 
must be identified, according to the compactification condition 
(\ref{COMPACT}). In other words, $c$ is also invariant under large gauge 
transformations. 

Summarizing, the zero mode $c$ is fully gauge
invariant. 

The remaining electromagnetic modes $A_m$ are decoupled from 
the theory due to the Gauss Law (\ref{GL}), as  will be shown below.

Next we consider  the 
expansion of the fermionic 
variables  in a background 
electromagnetic field. According to Ref. \cite{IsoMurayama}, these 
can be written as 
 
\begin{equation} 
\psi_1(x,t)=\sum_{n} a_n \phi_n(x){\rm e}^{-i\epsilon_n t}, \quad 
\psi_2(x,t)=\sum_{n} b_n \phi_n(x) 
{\rm e}^{i\epsilon_n t},\label{psi12} 
\end{equation} 
where  $a_n, b_n$  are standard fermionic  annihilation 
operators satisfying the non-zero anticommutators 
 
\begin{equation}\label{FACR} 
 \{ a_m, a_n^\dagger \} = \delta_{mn}, \quad \{ a_m, a_n \} = 0, \quad 
\{ b_m, b_n^\dagger \} = \delta_{mn}, \quad \{ b_m, b_n \} = 0, 
\end{equation} 
while any of the  $a$'s anticommutes with any of the  $b$'s. The states 
$\psi_1 \ (\psi_2)$ describe the positive \ (negative) chiral (eigenvalues of $\gamma_5$)    
sectors of 
the model. 
 
The basic functions $\phi_n$ , together with the 
eigenvalues of the energy are given by

\begin{equation} 
\phi_n(x)={1\over\sqrt L}e^{-i \epsilon_nx -ie\int_0^x 
A(z)dz},\quad 
\epsilon_n= 
{2\pi\over L}\left(n+{1\over2} -{eL\over 2\pi}c\right)\equiv {2\pi 
n\over L} + {\pi\over L}- ec. 
\label{VFP} 
\end{equation} 
 
Rewriting the fermionic sector  of the Hamiltonian density   
(\ref{HAMDEN}) as 
${\cal H}_F= 
\psi^\dagger h_F \psi$, we observe that  the corresponding   
eigenvalues of $h_F$ 
are 
$+ \epsilon_n$ and $- \epsilon_n$ for the positive and negative quirality 
sectors, respectively. 
 
Since $c$ is invariant under large and small gauge transformations the
energy eigenvalues $\epsilon_n$ are fully gauge invariant. Furthermore, according to the definition 
(\ref{VFP})
\be 
\phi_n \to {\rm e}^{ie\,\bar \alpha(x)}\phi_n,
\ee
under gauge transformations, where we have used $\bar \alpha(x)=0$ which is valid for both large and small gauge transformations.
As a consequence of the above
properties and in order to recover the transformation law (\ref{GGT}) of the fermionic field $\psi$, we must have 

\begin{equation}\label{GTAYB} 
a_n \rightarrow  a_n, \quad 
b_n \rightarrow b_n. 
\end{equation} 
for the gauge transformation of the fermionic operators $a_n$ and $b_n$. 
In other words, consistency among the compactification condition (\ref{COMPACT}), the transformation law (\ref{GGT}) and the definition (\ref{VFP}) demands that the basic 
fermionic operators $a_n$ and $b_n$ are {\bf fully gauge invariant in the compact case}.

Let us emphasize that the above property establishes a main difference  between the compact and the  non-compact cases. Following the same steps in
the latter situation we obtain that  the change  $c \rightarrow c - \frac{2 
\pi n}{e L }$, under large gauge   
transformations  and with these points not identified, implies that the energy eigenvalues are 
not gauge invariant, i.e.  $\epsilon_n \to \epsilon_{n+1}$, which leads to  
$\phi_n(x)  \to {\rm e}^{-ie \bar \alpha(x)} \phi_{n+1}$. Then, 
in order to satisfy the transformation property (\ref{GGT}) of the fermionic field, we must have now that $a_n\rightarrow a_{n+1}, b_n \rightarrow b_{n+1} $, under LGT.That is to say, the fermionic operators are not fully gauge invariant in the non-compact case. 

In this way, it is transparent that the topological behavior of $c$, i.e. compact versus non-compact variable, implies a completely different transformation law for the fermionic operators $a_{n}, b_n$ under gauge transformations.

Using the Fourier expansions (\ref{RFD}) and (\ref{psi12}), we rewrite the commutators for 
the fields in terms of the corresponding modes. 
In particular, the commutator $[ E(x), \psi_\alpha(y)]=0$ leads to 
\begin{equation}\label{CREa} 
[E_m, a_n] = \frac{ie}{2 \pi m }\left( a_n - a_{n+m}  \right), 
\ \  m \neq0, \qquad [E_0, a_n]=0, 
\end{equation} 
\begin{equation}\label{CREb} 
[E_m, b_n] = \frac{ie}{2 \pi m }\left( b_n - b_{n+m}  \right), 
\ \ m \neq0, \qquad [E_0, b_n]=0. 
\end{equation} 
The remaining commutators are 
\begin{equation}\label{RCOM} 
[ A_k, A_l]=0= [ E_k, E_l], \quad  [ A_k, E_l]= \frac{i}{L} \delta_{kl}, 
\quad [ A_k, a_n]=0, \quad [ A_k, b_m]=0. 
\end{equation}

Next, we concentrate on the commutator algebra of the Fourier modes. To this end, we
introduce the following  operators 
 
\begin{equation}\label{DC} 
  j_{++}^{nm}=a_n^\dagger a_m, \hspace{1cm} 
  j_{--}^{nm}=b_n^\dagger b_m, \hspace{1cm} 
  j_{+-}^{nm}=a_n^\dagger b_m, \hspace{1cm} 
  j_{-+}^{nm}=b_n^\dagger a_m, 
\end{equation} 
which satisfy $( j_{++}^{nm})^\dagger=  j_{++}^{mn}, \    
(j_{--}^{nm})^\dagger = 
j_{--}^{mn}$ \, and \, 
$(j_{+-}^{mn})^\dagger = j_{-+}^{nm}$. Additional useful combinations of the above fermionic operators are the currents
\begin{eqnarray}\label{JMMX} 
&&j_+(x)= \psi_1{}^\dagger(x) \psi_1(x)= \frac{1}{L} 
\sum_{n=-\infty}^{+ \infty} {\rm e}^{- \frac{2 \pi  i n}{L} x} 
\ j_+{}^n, \nonumber \\ 
&&j_-(x)= \psi_2{}^\dagger(x) \psi_2(x)= \frac{1}{L} 
\sum_{n=-\infty}^{+ \infty} {\rm e}^{+ \frac{2 \pi  i n}{L} x} \ 
j_-{}^n, 
\end{eqnarray} 
where 
\begin{equation}
\label{JMM} 
j_+{}^n =\sum_{m=-\infty}^{\infty} j_{++}^{m,m+n}, \quad j_+{}^0= Q_+, \qquad 
j_-{}^n =\sum_{m=-\infty}^{\infty} j_{--}^{m+n,m}, \quad j_-{}^0= Q_-. 
\end{equation} 
 
At this stage we introduce the $\zeta-$regularized form 
of the currents defined in Eq.(\ref{JMM}), 
 
\begin{equation} 
\label{REGJ} 
j_+{}^n|_{\rm reg}={\rm lim}_{s\rightarrow 0} 
\sum_{m=-\infty}^{\infty} \frac{1}{\lambda_{m,s}}a^\dagger_{m} a_{m+n}, \quad 
j_-{}^n|_{\rm reg}={\rm lim}_{s\rightarrow 0}\sum_{m=-\infty}^{\infty} 
\frac{1}{\lambda_{m,s}}b^\dagger_{m+n } b_{m}. 
\end{equation} 
 
In the sequel we will drop the subindex $ |_{\rm reg}$ from 
the above 
currents, but we will always consider their form (\ref{REGJ}) 
in calculating any relation involving 
them. At 
the end of the calculation we will take the $s\rightarrow 0$ 
limit. 
In other words, we will construct an algebra among regularized 
objects, which will be further restricted to the action upon the 
physical 
Hilbert space of the problem. 

As it is shown in detail the Appendix, the regularized current algebra
of the operators (\ref{REGJ}) is given by 
\begin{equation}\label{CRROFJ} 
 [ j_+{}^n,   (j_+{}^m)^\dagger ]= n \delta_{m,n}, \qquad 
[ j_-{}^n,  ( j_-{}^m)^\dagger ]= n \delta_{m,n}, \qquad  
[ j_+{}^n,   j_-{}^m ]=0. 
\end{equation} 
The  above commutation relations are the same as those obtained in the non-compact case.
The calculation 
here is somewhat different than the one in Ref. \cite{IsoMurayama}   
because we have used the regularized expressions (\ref{REGJ}) for the  currents. We have also verified in the Appendix that the regularized currents satisfy the hermiticity properties
\be
(j_\pm{}^m)^\dagger = j_\pm{}^{-m}. 
\ee

In order to satisfy the commutation relations (\ref{CREa}) 
and (\ref{CREb}) we make the ansatz 
\begin{equation}\label{AFORE} 
 E_m=\frac{1}{i L}\frac{\partial}{\partial A_m}  - \frac{e}{2 \pi  i m} 
 \left( j_+{}^m + (j_-{}^m)^\dagger \right),\ \ m\neq 0, \quad E_0= 
 \frac{1}{i L} \frac{\partial}{\partial c },  
\end{equation} 
which clearly satisfies the third commutation relation in (\ref{RCOM}). 
Let us emphasize  that the fields $E_m, \ m \neq 0$, defined in 
Eq.(\ref{AFORE}) satisfy 
also  the second commutation relation  in (\ref{RCOM}), 
by virtue of (\ref{CRROFJ}). Substituting the  expressions 
(\ref{AFORE}) 
in the corresponding commutators of Eq.(\ref{CREa}) we obtain 
\begin{equation}\label{DAODA} 
\frac{\partial a_n}{\partial A_m}= - \frac{e L}{ 2 \pi m} a_n, \ 
m \neq 0,\quad \frac{\partial a_n}{\partial c}=0. 
\end{equation} 
The above equation leads to the following solution for the fermionic 
operators with respect to their dependence on the gauge field 
\begin{equation}\label{SOLFO} 
 a_m= {\rm exp} \left( -\frac{e L}{2 \pi } 
\sum_{k\neq0} \frac{1}{k} A_k \right) {\bar a}_m \equiv U {\bar 
a}_m, 
\end{equation} 
where ${\bar a}_m$ are new fermionic operators which are independent 
of the gauge field $A_k$ 
and which also  satisfy the basic fermionic  anticommutation relations. 
In fact, the transformation $U$ defined above is unitary because 
$ A_k{}^\dagger= A_{-k}$. A relation analogous to (\ref{SOLFO}) can be 
found for the remaining  fermionic operators:  $b_m =U {\bar b}_m$, 
where the new operators $\bar b_m$ are independent of the 
electromagnetic potential by construction.

The  expression (\ref{SOLFO}) reproduces the fully gauge invariant character of $a_n$. In fact, under the gauge transformation given by (\ref{GTOM}), 
the exponential in (\ref{SOLFO})  changes by a factor 
${\rm exp}( i e \sum_{k \neq 0} {\bar \alpha}_k) $ which is exactly 
${\rm exp}( i e {\bar \alpha}(0))=1$, according to the relation 
(\ref{ALFA0}). The same result is obtained for the operators $b_n$.
 
With the ansatz (\ref{AFORE}), the Gauss law 
${\cal G}(x) =- \frac{1}{L}\sum {\rm exp} (-\frac{2 \pi i mx}{L}) 
{\cal G}_m $ reduces to 
\begin{eqnarray} 
\label{GAUSSL} 
{\cal G}_0&=& e \left( j_+{}^0 + (j_-{}^0)^\dagger \right) =e Q 
\nonumber \\ 
{\cal G}_m &=& 2 \pi i m \ E_m + e \left( j_+{}^m + 
(j_-{}^m)^\dagger \right)=\frac{2 \pi m}{L} \frac{\partial}{\partial A_m}, \quad 
m\neq 0. 
\end{eqnarray} 
These expressions reproduce the transformation property 
\begin{equation}\label{TPGL} 
[{\cal G}(x), a_n] = e a_n \delta(x) \Longleftrightarrow 
[  {\cal G}_m , a_n] = -e \ a_n, 
\end{equation} 
in virtue of the relation (\ref{DAODA}).  The above commutator is the 
infinitesimal version 
of the relation (\ref{GTAYB}). 
   
Our expression  (\ref{GAUSSL}) for the Gauss law constraint 
is somewhat different from the one obtained in  Ref. \cite{IsoMurayama}. 
Imposing this constraint upon the physical states 
we conclude that the wave functions of the system must 
be of zero electric  charge and also independent of the modes $A_m, \ m \neq0$, which explicitly show the decoupling of the $m\neq0$ electromagnetic modes.

In particular, the following bilinears in the fermionic operators: $a^\dagger  a,\, a^\dagger  b, \, b^\dagger a, \, b^\dagger b $, which will be subsequently used in the solution of the model,  are independent of the electromagnetic modes and are gauge invariant under SGT and LGT. 
  
Summarizing, we have shown that the compactification condition (\ref{COMPACT}) implies that  the operators $c$, $a_n$ and $b_n$ are fully gauge invariant.  Also, the Gauss constraint imply that
the wave function of the system is independent of the electromagnetic modes $A_m,\ m\neq 0$.

\section{Fock Space}

We now construct the fermionic 
Fock space in a background electromagnetic field. Starting from the   
vacuum 
$|0 \rangle$ annihilated by the operators $a_n$ (positive quirality   
sector), 
$b_n$  (negative quirality sector), 
the Dirac vacuum $|vac \rangle$ 
is constructed  in such a way that all negative energy levels are 
filled. Our compactification condition  (\ref{COMPACT}) for the   
electromagnetic variable $c$ implies that  all  levels with $n \leq -1    
\ ( n\geq 0 
) $ have negative energies for the 
 positive  \ (negative) quirality sectors, respectively.  In this   
way, the 
Dirac vacuum is 
\begin{equation}\label{DVD} 
  |vac \rangle = 
                 \prod_{n=-\infty}^{-1}a_n^\dagger |0 \rangle \otimes 
                 \prod_{n=0}^{\infty}b_n^\dagger |0 \rangle. 
\end{equation}

The $\zeta-$regularized Hamiltonians and the  $\zeta-$regularized 
charge operators of the 
positive (negative) chiral sectors are given by \cite{IsoMurayama} 
 
\begin{eqnarray} 
  H_+ &=& \lim_{s \rightarrow 0} \sum_{n\in Z} 
\frac{\epsilon_n}{\lambda_{n,s}} a_n^\dagger a_n 
{\hskip 1cm} 
  Q_+= \lim_{s \rightarrow 0} \sum_{n\in Z}\frac{1}{\lambda_{n,s}} 
 a_n^\dagger a_n \nonumber \\ 
H_- &=& \lim_{s \rightarrow 0} \sum_{n \in Z} 
\frac{(-\epsilon_n)}{\lambda_{n,s}} 
b_n^\dagger b_n {\hskip 1cm} 
Q_-= \lim_{s \rightarrow 0} \sum_{n\in Z}\frac{1}{\lambda_{n,s}} 
b_n^\dagger b_n, 
\label{DOB} 
\end{eqnarray} 
where the regulator is given by $\lambda_{n,s}= |\lambda\, 
\epsilon_n|^s$, with 
$\lambda$ been a parameter with dimensions 
of inverse energy. 
 
The total charge $Q$, the total chiral charge $Q_5$ and the total 
energy $H$ are hermitian operators defined by 
 
\begin{equation}
\label{DEFCH} 
  Q=Q_+ + Q_-, \quad Q_5= Q_+ -Q_-, \quad H_F= H_+ + H_-. 
\end{equation} 
The  eigenvalues of the above operators for the Dirac vacuum state are 
 
\begin{eqnarray} 
  Q |vac\rangle &=& 0, 
\qquad Q_5 |vac\rangle = -\frac{ecL}{\pi } |vac \rangle , 
\nonumber \\ 
  H_F |vac\rangle &=& \frac{2\pi}{L}\left \{ \left( \frac{ecL}{2\pi}\right)^2 
-\frac{1}{12} \right \} 
 |vac\rangle \equiv {\varepsilon}_0 |vac\rangle, 
\label{EDV} 
\end{eqnarray} 
At this level it is already convenient to introduce the 
modified chiral charge 
\begin{equation} 
\label{MCC} 
{\bar Q}_5= lim_{s\rightarrow0}\sum_{n= -\infty}^{+ \infty} 
 \frac{1}{|\lambda \epsilon_n|^s}\left( { a}_n^\dagger { a}_n 
- { b}_n^\dagger { b}_n  \right)+\frac{ecL}{\pi}, 
\end{equation} 
which is invariant under SGT and LGT 
in the compact Schwinger model. This is not the case in the non-compact 
situation, where ${\bar Q}_5$ is not invariant under LGT. The eigenvalues of ${\bar Q}_5$ are integer 
numbers in the fermionic Hilbert space. In the sequel we will refer 
to ${\bar Q}_5$ as the (total) modified chiral charge of the system. Also, a given eigenvalue $2\, M$ of  
${\bar Q}_5$ will be referred to as the $M$-chiral sector of the theory. In this way, the 
second equation (\ref{EDV}) reads ${\bar Q}_5 |vac\rangle=0$, which 
assigns zero chiral charge to the Dirac vacuum.

In order to construct  excited states with different modified chiral charge, we need to use the operators
(\ref{DC}). The application of any of them  to  $| vac \rangle$   
produce new 
energy eigenstates  with the same zero eigenvalue for the electric   
charge. 
Nevertheless, the chiral charge is changed in steps of two units, by the action of   $j_{+-}$, 
$j_{-+}$. 
 
Let us focus on the states 
\begin{equation}\label{DCCS} 
  j_{+-}^{pq} |vac \rangle, \ p,q \geq 0 \hspace{1cm} j_{-+}^{pq}   
|vac \rangle, 
\ p,q < 0, 
\end{equation} 
where the choice of $p,q$ in each case is such that  a non-zero   
vector results. 
If we compute their energy, we find 
\begin{eqnarray} 
   H_F(j_{+-}^{pq} |vac \rangle )&=& 
                    ({\varepsilon}_0+\epsilon_p +\epsilon_q )j_{+-}^{pq} |vac 
\rangle , 
 \nonumber \\ 
   H_F(j_{-+}^{pq} |vac \rangle )&=& 
             ( {\varepsilon}_0-\epsilon_{p} -\epsilon_{q} )j_{-+}^{pq} |vac 
\rangle . 
 \label{SEDC1} 
\end{eqnarray} 
 
The chiral charge of these states is 
\begin{eqnarray} 
   {\bar Q}_5(j_{+-}^{pq} |vac \rangle )&=& 
               \ 2 j_{+-}^{pq} |vac \rangle, 
 \nonumber \\ 
   {\bar Q}_5(j_{-+}^{pq} |vac \rangle )&=& 
               -2 j_{-+}^{pq} |vac \rangle. 
 \label{SEDC2} 
\end{eqnarray} 
 From Eqs.  (\ref{SEDC1}) and (\ref{SEDC2}) we  conclude that the 
state with chiral charge 
$+ 2$ and lowest energy ${\varepsilon }_1={\varepsilon}_0+2\epsilon_0$  is 
\begin{equation} 
  j_{+-}^{00}|vac \rangle = 
                 \prod_{n=-\infty}^0 a_n^\dagger 
                 \prod_{n=1}^{\infty}b_n^\dagger |0 \rangle 
                    \equiv \left| {\varepsilon}_1, 2 \right \rangle. 
\end{equation} 
Analogously,  the 
state with chiral charge 
$-2$ and lowest energy ${\varepsilon}_{-1}={\varepsilon}_0-2\epsilon_{-1}$ 
is 
\begin{equation} 
  j_{-+}^{-1-1}|vac \rangle = 
                 \prod_{n=-\infty}^{-2} a_n^\dagger 
                 \prod_{n=-1}^{\infty}b_n^\dagger |0 \rangle 
                  \equiv \left|{\varepsilon }_{-1}, -2 \right. \rangle. 
\end{equation} 
Repeating this procedure we construct the local ground states $\left |{\varepsilon}_N, 2N \right \rangle$, for a definite quirality $2\,N, \,  \ N \in {\cal Z} $,  
 
\begin{eqnarray}  
 j_{+-}^{NN}\left |{\varepsilon}_N, \, 2N \right \rangle &=& 
                     \left | {\varepsilon}_{N+1},\, 2(N+1) \right \rangle, 
\nonumber \\ 
 j_{-+}^{N-1N-1}\left |{\varepsilon}_N,\, 2N \right \rangle &=& 
                     \left |{\varepsilon}_{N-1},\, 2(N-1) \right \rangle, 
\label{COV} 
\end{eqnarray} 
 having minimum energy ${\varepsilon}_N,$ determined by the recursions ${\varepsilon}_{N+1}={\varepsilon}_{N}+2\epsilon_{N}$ and ${\varepsilon}_{N-1}={\varepsilon}_{N}-2\epsilon_{N-1}$, respectively. The above   
states can be written as 
\begin{equation}\label{VED} 
  \left |{\varepsilon}_N,2N \right \rangle = 
                 \prod_{n=-\infty}^{N-1}a_n^\dagger |0 \rangle \otimes 
                 \prod_{m=N}^{\infty}b_m^\dagger |0 \rangle. 
\end{equation} 
Using the regularized expression for the fermionic Hamiltonian, the result 
\begin{equation} 
\label{VACEN} 
{\varepsilon}_N(c)=\frac{2 \pi}{L}\left\{\left( N- \frac{ecL}{2 \pi} \right)^2-\frac{1}{12}\right\}, 
\end{equation} 
is obtained, which is equivalent to what we get using  the recursions previously indicated.  
In the above notation the Dirac vacuum state is 
\begin{equation}\label{V} 
   |vac \rangle = \left | {\varepsilon}_0, 0 \right \rangle. 
\end{equation} 
All the states (\ref{VED}) have zero electric charge, which is not 
explicitly written. 
 
Summarizing, from the Dirac  vacuum we have so far  constructed states with 
minimum energy for each 
possible chirality. As noticed before, each one of these states can be considered as a 
{\it local vacuum} in the corresponding  quirality sector. 
 
We  are now in position to determine the characteristics of the complete fermionic spectrum in the electromagnetic 
external field. Within each chiral sector and starting from the local ground state $|{\varepsilon}_N, 2N \rangle$, we have to consider all possible 
zero-charge excitations generated by  $a^\dagger_p a_q $. The corresponding minimum 
energy ${\varepsilon}_N$ will be  shifted to ${\varepsilon}_N + \epsilon_p - \epsilon_q= 
{\varepsilon}_N +\frac{2 \pi}{L}(p-q)$. Furthermore, the expression (\ref{VED}) implies that 
\begin{equation} 
\label{VACCOND1} 
a^\dagger_p a_q |{\varepsilon}_N, 2N \rangle=0, \quad p < q, \quad q> N-1. 
\end{equation}   
In this way,  only the excitations with $p \geq q$ are allowed. An analogous result is obtained by considering the excitations generated by the operators $b_n^\dagger b_m$. The final conclusion is that the fermionic spectrum is 
\be
\{ {\varepsilon}_N + \frac{2 \pi}{L} M, M=1,2, \dots, ,\ N=0, \pm1, \pm 2, \dots  \}.
\ee 
 
Next we  explicitly construct the  corresponding excited states. To this end 
we use the operators (\ref{JMM}). The electric charge $Q$ and  the modified chiral 
charge ${\bar Q}_5$ of the system  are thus written as 
\begin{equation} 
Q= j_+{}^0 +j_-{}^0, \qquad {\bar Q}_5=j_+{}^0 -j_-{}^0 +\frac{e c   
L}{\pi}. 
\end{equation}

The operators $j_\pm{}^n, n\geq 1$  annihilate the states 
(\ref{VED}), i.e., 
\begin{equation} 
\label{JPMV} 
j_\pm{}^n \left |{\varepsilon}_N,2N \right \rangle =0,\quad n\geq 1,  
\end{equation} 
which is just a consequence of the relation (\ref{VACCOND1}), together with the analogous 
one including the $b$-operators. Another property that will be used in the sequel is 
\begin{equation} 
\label{CHVESP} 
\langle {\varepsilon}_N,2N \, |\, j_{+ -}^{p\,q} \, |{\varepsilon}_N,2N  \rangle=0,\quad  
\langle {\varepsilon}_N,2N \, |\, j_{- +}^{p\,q} \, |{\varepsilon}_N,2N
\rangle=0, \quad \forall \ p,\, q. 
\end{equation} 
 
Also, the following commutators can be calculated 
\begin{equation} 
\label{CHFJ} 
\left[H_F, (j_{\pm}^n)^\dagger \right]= \frac{2 \pi n}{L} \ (j_{\pm}^n)^\dagger. 
\end{equation} 
 
In this way, the fermionic Fock space in the background electromagnetic field   
will consist 
of all the local vacuums  (\ref{VED}), together with all  possible 
states constructed from them by the application of an arbitrary number 
of the current operators ${(j_\pm{}^n })^\dagger, n=1,2, \dots$ defined in Eq. (\ref{REGJ}).

Taking into account the spectrum of the system, together with the way in which the Fock space 
has been constructed, the fermionic Hamiltonian in the external field can be rewritten in the Sugawara form \cite{SUGA}
\begin{equation} 
\label{HAMCUAD} 
H_F= {\varepsilon}_N(c)+ \frac{2\pi}{L} \sum_{n>0} \left((j_+^n)^\dagger  j_+^n + (j_-^n)^\dagger  j_-^n\right). 
\end{equation} 
 
Also we obtain the following commutation relations 
 
\begin{equation}\label{RCG} 
[ j_{+}^n,j_{+-}^{pq}]=j_{+-}^{p-n,q}, \quad 
[ j_{-}^n,j_{+-}^{pq}]=-j_{+-}^{p,q-n}, \quad 
[ j_{+}^n,j_{-+}^{pq}]=-j_{-+}^{p,q+n}, \quad 
[ j_{-}^n,j_{-+}^{pq}]=j_{-+}^{p+n,q} 
\end{equation} 
From them, together with (\ref{CRROFJ}) we 
conclude that 
\begin{equation}\label{ONCC} 
   [ Q,j_{+}^n]=[ {\bar Q}_5,j_{+}^n]=[ Q,j_{-}^n]=[ {\bar   
Q}_5,j_{-}^n]=0, 
\end{equation} 
and 
\begin{equation}\label{OSCC} 
[ Q,j_{+-}^{pq}]=[ Q,j_{-+}^{pq}]=0, \hspace{.5cm} 
[ {\bar Q}_5,j_{+-}^{pq}]=2j_{+-}^{pq}, \hspace{.5cm} 
[ {\bar Q}_5,j_{-+}^{pq}]=-2j_{-+}^{pq}. 
\end{equation} 
The above commutators show that the current operators $j_+^n$ and   
$j_-^n$ do 
not change either the electric or  the chiral charge. Also we   
conclude that any 
linear 
combination of the operators $j_{+-}^{pq}\,\, (j_{-+}^{pq})$, although does not change 
the electric charge, will increase (decrease) the modified chiral charge by 2 units.

\section{THE FULL QUANTIZATION  OF THE  MODEL}

The next step  is to write  the complete  Hamiltonian 
$H=H_{EM}+H_{F}$ in terms of the fermionic currents operators 
together with 
the electromagnetic degrees of freedom, which are   the zero 
mode of the electric 
field, $\partial /\partial c$, and the zero  mode of the gauge potential, 
$c$, 
\begin{eqnarray}\label{FH} 
 H_{EM}= \frac{L}{2}\sum_n E_n^\dagger E_n&=&-\frac{1}{2L} \left ( \left ( \frac{\partial}{\partial c} 
\right )^2 -\sum_{n\not= 0} \left ( \frac{eL}{2\pi n}\right )^2 
        (j_+^n+(j_-^n)^\dagger ) ((j_+^n)^\dagger + j_-^n)\right ), 
\nonumber  \\ 
H_F&=&\frac{2\pi}{L} \left ( \frac{Q_+^2+Q_-^2}{2}-\frac{1}{12}+ 
    \sum_{n>0} ((j_+^n)^\dagger  j_+^n + (j_-^n)^\dagger  j_-^n)\right ). 
\end{eqnarray} 
Following Ref. \cite{Link,IsoMurayama}, we have explicitly used  the   
Gauss law constraint (\ref{GAUSSL}) to express the electric field modes $E_m$ in terms of the fermionic currents.

We  can further split the total Hamiltonian  into  a  zero-mode   
part, which 
decouples from the fermionic sector, obtaining 
\begin{equation}\label{HM} 
 H=H_{EM}+H_{F}=H_0+\sum _{n>0}H_n-\frac{2\pi}{12 L}, 
\end{equation} 
where 
 \begin{eqnarray}\label{HME} 
  H_0&=&\frac{\pi}{2 L} \left ( Q^2+\left({\bar Q}_5- 
\frac{e c L}{\pi}\right)^2 \right) -\frac{1}{2L} 
 \left(\frac{\partial}{\partial c} \right)^2, 
\nonumber \\ 
  H_n&=& \frac{2\pi}{L} 
  ((j_+^n)^\dagger  j_+^n + (j_-^n)^\dagger  j_-^n ) + 
  \frac{e^2L}{4 \pi^2 n^2} 
  ((j_+^n)^\dagger + j_-^n) (j_+^n+(j_-^n)^\dagger). 
\end{eqnarray} 
 
Even though at this stage $c$ is a dynamical variable such that   
$\dot c \neq 
0$, the  modified chiral 
charge ${\bar Q}_5$ is conserved in the full Hilbert space of the   
problem, as will be shown in more detail in section VII. 
 
In order to diagonalize the expression (\ref{HME})  for the   
Hamiltonian, it is 
convenient  to introduce the Bogoliubov transformations for the currents 
operators,  given by \cite{Link,IsoMurayama} 
\begin{eqnarray} 
\label{BOGT} 
 &&{\tilde j}_+ ^n= U_n^\dagger\,(j_+ ^n )\,U_n = {\rm cosh}t_n\,{ j}_+ ^n - 
 {\rm sinh}t_n\,({ j}_- ^n)^\dagger,\nonumber \\
&& ({\tilde j}_- ^n)^\dagger = U_n^\dagger\,(j_- ^n )^\dagger \,U_n =
-{\rm sinh}t_n\,{ j}_+ ^n + 
 {\rm cosh}t_n\,({ j}_- ^n)^\dagger, 
\end{eqnarray} 
with
\begin{eqnarray}
&& {\rm cosh} 2t_n= \frac{1}{{\cal E}_n}\left(\frac{2\,\pi\,n}{L} + \frac{e^2\, L}{4\, \pi^2\, n}\right), \quad {\rm sinh} 2t_n=\frac{1}{{\cal E}_n}\,\frac{e^2\, L}{4\, \pi^2\, n}, \quad
{\cal E}_n=\sqrt{\left(\frac{2\,\pi\,n}{L} \right)^2 + \frac{e^2}{\pi}}.  
\end{eqnarray}
The unitary operators $U_n$ which produce the above transformations are 
\begin{equation} 
U_n=\exp \left\{ -\frac{t_n}{n}\left(({j_+}{}^n)^\dagger 
({j_-}{}^n)^\dagger -  {j_+}{}^n {j_-}{}^n\right) \right\}. 
\end{equation} 
The complete transformation is obtained  through the operator $U= \prod_{n\geq 1}U_n   
 $.

In particular,  we have  that 
\begin{equation} 
\label{COMJ0U} 
[ Q,   U_n ]= [ {\bar Q}_5,  U_n   ] = 0, \quad n \geq 1. 
\end{equation} 
Summarizing, the Bogoliubov transformation 
affects  only  the  fermionic  modes of the system and, in particular, the 
currents $j_{\pm}^0$, or equivalently $Q$ and ${\bar Q}_5$,  remain unchanged.

In this way,  the fully  rotated  Hamiltonian 
\begin{equation}
{ H}_B=U^\dagger \,H( { j}_+^n, { j}_-^n) \, U= H( { {\tilde j}}_+^n, { {\tilde j}}_-^n), 
\end{equation}
 is 
 \begin{eqnarray} 
\label{HTOTAL} 
 { H}_B= \frac{\pi}{2L} \left(Q^2+\left({\bar Q}_5- 
\frac{e c L}{\pi}\right)^2 \right) 
      -\frac{1}{2L} \left(\frac{\partial}{\partial c} \right)^2 
+ \sum_{n>0} \frac{{\cal E}_n}{n}\left(({ j}_+^n)^\dagger 
{ j}_+^n + ({ j}_-^n)^\dagger { j}_-^n \right), 
\end{eqnarray} 
up to an infinite constant. 

To construct the 
Hilbert space of the full theory we  start from the states
$ \left| {\varepsilon}_N,2N \right\rangle $ given in Eq. (\ref{VED}), which have  minimum energy, zero electric charge and  eigenvalues
$2\, N$ for  the modified chiral charge ${\bar Q}_5$. 
In particular, we recall that the states (\ref{VED})  
are  annihilated by the 
currents ${ j}_+{}^{n}$ and ${ j}_-{}^{n}, \ 
n \geq 1 $, for all $N$.

As in the non-compact case, Eq. (\ref{HTOTAL}) implies that each mode decouples, i.e. the total 
energy ${\cal E}$ is  
given by the superposition of each energy mode. 
Also, the total wave function $\Delta$, such that $H \Delta= {\cal E} 
\Delta $, 
is  
$\Delta= \Pi_n \Delta_n$,  where  $ \Delta_n$ are the eigenfunctions of $H_{B\, n}$.
 
The loop 
approach, which 
we are taking as a guide in defining the compact Schwinger model, 
requires 
only the compactification of the zero mode $n=0$, which is related 
to the 
electromagnetic field. Nevertheless, since the $n\geq 1$ modes will be obtained by applying
raising operators to the zero modes, the compactification process propagates to the excited states, leading also to different spectra and eigenfunctions,  when compared with the non-compact case. 
 
The general structure of the states in the Hilbert space of  the model will be 
of the type 
\begin{equation} 
\label{GENFORM} 
|\hbox{state} \rangle= F(c) \times |\hbox{fermionic}\rangle. 
\end{equation} 
The whole wave function will have zero electric charge and definite   
chiral charge, which 
really implies a condition only upon de fermionic 
piece. The  strategy to construct the Hilbert space  will be to start 
from the zero modes $ F_N(c) \times {\left|N \right   
\rangle}$   and to subsequently apply  
all possible combinations of the 
raising operators $({ j}_\pm{}^m)^\dagger$. In this way, we will 
obtain an infinite tower of states for each value of the modified chiral charge.

\subsection{The zero modes}

They correspond to the case of zero fermionic excitations and can be written as
\begin{equation} 
\label{ZERMOD} 
{ |N \rangle}_B= F_N(c) \times \left| {\cal E}_N,2N \right\rangle. 
\end{equation}
The subscript $B$ in any ket is to remind us that such vector is written in the Bogoliubov rotated
frame, where the Hamiltonian has the form (\ref{HTOTAL}).  
Its action  upon the above wave functions reduces to the 
following Schroedinger equation for the zero mode wave functions 
$F_N(c)$ 
\begin{equation} 
\label{ZEROM} 
\left(-\frac{1}{2L}\left( \frac{\partial}{\partial c} \right )^2 
+ \frac{e^2L}{2\pi}\left( \frac{2\pi N}{eL} -c \right )^ 2 
\right)F_N(c)= {\cal E}_{N,0}\, F_N(c) 
\end{equation} 
 
A fundamental difference  between the compact and the 
non-compact model arises in the energy spectrum $\{{\cal E}_{\alpha, N, 0}\}$ of  zero 
mode sector. Here $\alpha=0,1,2, \dots $ labels the eigenvalues of the zero mode $0$ in the $N$-chiral  sector of the model. 
The wave functions $F_N(c)$ satisfy a Schroedinger 
equation corresponding to a  piecewise harmonic oscillator  
described  by the potentials $V_N(c)$ shown in Fig.1. 
Each of
these potentials is defined in the interval $-{\bar c}\leq c \leq   
{\bar c},$ with ${\bar c}=\frac{\pi}{e L}$.  

For arbitrary functions $F(c)$ and $G(c)$ in our Hilbert space,  we define their inner product in the standard way
\begin{equation}
\label{scalarp}
\langle F | G \rangle = \int^{\bar c}_{-\bar c} dc\ F^*(c) G(c).
\end{equation}
In order to determine the appropriate boundary conditions we demand the hermiticity of the zero mode electric field operator $E_0=\frac{1}{i\,L}\,
\frac{\partial}{\partial\,c}$. This leads to require 
\begin{equation} 
\label{BCWF} 
F_N|_{c=-{\bar c}}=F_N|_{c=+{\bar c}},
\ee
for the wave function $F_N$. Furthermore, hermiticity of the Hamiltonian (\ref{ZEROM}) implies the  additional
boundary condition
\be
\label{BCDWF} 
{\partial F_N \over \partial c}|_{c=- {\bar c}}= 
{\partial F_N \over \partial c}|_{c=+ {\bar c}}, 
\end{equation} 
for the derivative of the wave function.

In this way, our boundary conditions (\ref{BCWF}) and (\ref{BCDWF}) are an unavoidable  consequence of the compactification of the electromagnetic degree of freedom $c$, together with standard hermicity requirements. They are completely analogous 
to those of the one-dimensional rigid rotor, which provide both the correct eigenvalues for the z-component of the angular momentum operator and the correct energy spectrum.

The above boundary conditions should be contrasted with those used in  Eqs. (3.15) of Ref. \cite{Manton}, and  Eq. (48) of Ref.
\cite{Link}, for example. The latter
are correctly designed to recover
the non-compact case, i.e. to unfold the circle into the line. Moreover, for a given $L$, they cannot be 
continuously related to those in Eqs.(\ref{BCWF}) and (\ref{BCDWF}). This emphasizes 
the non equivalence of both models, which has its origin a different choice of topology via   the compactification condition (\ref{COMPACT}). 
 
The above Schroedinger equation (\ref{ZEROM}) together with the boundary conditions 
(\ref{BCWF}) and (\ref{BCDWF}) lead to 
energies which are not any more given by the characteristic equally   
spaced 
harmonic oscillator 
spectrum, as it is the case in  the non-compact model. 
 
The solution corresponding to $N=0$ has been already discussed in   
Ref.\cite{GMVU}, 
together with the corresponding wave functions. Here we extend the   
calculation 
for arbitrary $N\neq 0$. To this end, let us introduce the auxiliary   
variables 
\begin{equation}\label{AVE} 
 y= \sqrt{ \frac{2eL}{\sqrt \pi}}\left ( \frac{2N\pi}{eL}-c \right ), 
 \hspace{2cm} 
{\cal E}_{\alpha, N,0}=-\frac{e}{\sqrt \pi}\, a_{\alpha, N},  
\end{equation} 
where $y$ and $a_N$ are dimensionless quantities. The range of $y$ is  
\begin{equation} 
  y_{N_-}\leq y \leq y_{N_+}, \hspace{0.5cm} 
  y_{N_-}\equiv \sqrt{ \frac{2\sqrt {\pi ^3}}{eL}} (2N-1), \hspace{0.5cm} 
  y_{N_+}\equiv \sqrt{ \frac{2\sqrt {\pi ^3}}{eL}} (2N+1) 
\end{equation} 
and the Eq. (\ref{ZEROM}) reduces to  
\begin{equation}\label{FDE} 
  f''- \left( \frac{1}{4}y^2+a_N\right) f=0, 
\end{equation} 
where $F_N(c)= f(\frac{2 \pi N}{e L}-c)$. 
The  boundary conditions Eqs.(\ref{BCWF}) and (\ref{BCDWF}) on $f(y)$ are 
\begin{equation} 
\label{BCY} 
  f(y_{N_-})=f(y_{N_+}), \hspace{2cm} f'(y_{N_-})=f'(y_{N_+}). 
\end{equation} 
 
The general solution of   Eq. (\ref{FDE}) is expressed in terms  
of cylindrical parabolic functions \cite{Abramowitz} 
\begin{equation}\label{SOLU} 
  f(y)=A e^{-y^2/4} 
  M\left( \frac{a}{2}+\frac{1}{4},\frac{1}{2},\frac{y^2}{2} \right ) 
      +Bye^{-y^2/4} 
  M\left( \frac{a}{2}+\frac{3}{4},\frac{3}{2},\frac{y^2}{2} \right ), 
\end{equation} 
where $M(b,c,z)$ is the confluent hypergeometric function.   
 It will be convenient introduce the new dimensionless label 
\begin{equation} 
 l=\frac{eL}{\pi^{3/2}}. 
\end{equation} 
The eigenvalue conditions for Eq.(\ref{FDE}) will determine the energy  levels 
${\cal E}_{\alpha, N, 0}(a)$ as a function of $l$.  From the boundary conditions (\ref{BCY}) we obtain  ($N \neq 0$)   
\begin{eqnarray}\label{CFEN} 
& &\left[e^{-y_{N_-}^2} 
M\left( \frac{a}{2}+\frac{1}{4},\frac{1}{2},\frac{y_{N_-}^2}{2} \right ) 
 -e^{-y_{N_+}^2} 
M\left( \frac{a}{2}+\frac{1}{4},\frac{1}{2},\frac{y_{N_+}^2}{2} \right ) 
\right] \times \nonumber \\ 
& & \left[e^{-y_{N_-}^2}\left( \left(1-\frac{y_{N_-}^2}{2}\right) 
M\left( \frac{a}{2}+\frac{3}{4},\frac{3}{2},\frac{y_{N_-}^2}{2} \right ) 
+y_{N_-}^2\left(\frac{a}{3}+\frac{1}{2}\right) 
M\left( \frac{a}{2}+\frac{7}{4},\frac{5}{2},\frac{y_{N_-}^2}{2} \right ) 
\right) \nonumber  \right.\\  
&-&\left. e^{-y_{N_+}^2}\left( \left(1-\frac{y_{N_+}^2}{2}\right) 
M\left( \frac{a}{2}+\frac{3}{4},\frac{3}{2},\frac{y_{N_+}^2}{2} \right ) 
+y_{N_+}^2\left(\frac{a}{3}+\frac{1}{2}\right) 
M\left( \frac{a}{2}+\frac{7}{4},\frac{5}{2},\frac{y_{N_+}^2}{2} \right ) 
\right) \right] \nonumber \\ 
&=&\left[e^{-y_{N_-}^2} 
M\left( \frac{a}{2}+\frac{3}{4},\frac{3}{2},\frac{y_{N_-}^2}{2} \right ) 
 -e^{-y_{N_+}^2} 
M\left( \frac{a}{2}+\frac{3}{4},\frac{3}{2},\frac{y_{N_+}^2}{2} \right ) 
\right ] \times \nonumber \\ 
& & \left[e^{-y_{N_-}^2}\left(-\frac{y_{N_-}^2}{2} 
M\left( \frac{a}{2}+\frac{1}{4},\frac{1}{2},\frac{y_{N_-}^2}{2} \right ) 
+y_{N_-}^2\left(a+\frac{1}{2}\right) 
M\left( \frac{a}{2}+\frac{5}{4},\frac{3}{2},\frac{y_{N_-}^2}{2} \right ) 
\right) \right. \nonumber \\ 
& &\left. -e^{-y_{N_+}^2}\left(-\frac{y_{N_+}^2}{2} 
M\left( \frac{a}{2}+\frac{1}{4},\frac{1}{2},\frac{y_{N_+}^2}{2} \right ) 
+y_{N_+}^2\left(a+\frac{1}{2}\right) 
M\left( \frac{a}{2}+\frac{5}{4},\frac{3}{2},\frac{y_{N_+}^2}{2} \right ) 
\right) \right], 
\end{eqnarray} 
which defines the function $a_{\alpha, N}=a_{\alpha, N}(l)$.  
 
As in the $N=0$ case \cite{GMVU},  
this function can  
only be determined numerically for arbitrary $l$. 
In Fig.2 and Fig. 3  we show 
the results for $a_{\alpha, N}$ versus $l$, for the choices $\alpha = 0,1,2$ and $N=0, 1, 2, 3$. 
 
Some properties of the above quantization condition, together with  its solutions are the  following: 
 
(1) As can be seen from Fig.1, there is the symmetry $V_n(c)= V_{-n}(-c)$ among the potentials $V_N(c)$. Thus, the values of $a_N$ are the same in  
both cases. This can be seen from the fact that when  
$n \rightarrow -n$,  
$y_{N_-}^2 \rightarrow y_{N_+}^2$, 
$y_{N_+}^2 \rightarrow y_{N_-}^2$ leading to the invariance of   
equation (\ref{CFEN}). 
  
(2) From the numerical calculation we find that $a_{0, N}$ is  
monotonously increasing and also that  
$\lim _{l\rightarrow \infty} a_{0, N}(l)=0$. This last property is  
consistent with the fact that if $a_{0, N}$ remains finite when 
$l\rightarrow \infty$, then $a_{0, N}=0$. 
 
(3) The behavior for negative $a_{\alpha, N}$  and large  
absolute value 
$(|a_{\alpha, N}|\gg 1)$, when $l\gg 1$, is given by  
\begin{equation} 
 a_{\alpha, N}(l)=-\frac{\pi^2}{2}\left( 2 \left[ \frac{\alpha}{2} 
\right ] -1 \right ) ^2l, \hspace{2cm} \alpha=1,2, \dots 
\end{equation} 
where $[x]$ is the maximum integer function. 
  
(4) The behavior of $a_{\alpha, N}(l)$ for $l\rightarrow 0$ is given by the  
limit 
\begin{equation} 
  \lim_{l\rightarrow 0} 
  \frac{1}{\Gamma (\frac{a}{2}+\frac{1}{4}) 
                \Gamma (\frac{a}{2}+\frac{1}{4})} 
  e ^{\frac{(2N+1)^2}{l}}l^{1-a} =0  \Rightarrow a \rightarrow - \infty. 
\end{equation} 
 
Physically these limits are understood in the following way. Since  the potential function at the end-points $c=\pm {\bar c}$ 
behave as $1/L$ for each $N$ (Fig. 1),  
the corresponding  values  go to  
zero and the interval of definition of $c$ is very small,  when $L \rightarrow \infty$. Then,  
for every $N$,  all potential functions look like a one-dimensional box with infinite walls. Thus, in this limit the energy  
eigenvalues correspond to the  one-dimensional rotator. 
 
In analogous manner, we can understand the behavior for  
$L \rightarrow 0$. In this case  
$V(c) \rightarrow \infty$ for all $N\neq 0$, while the domain of  
$c$ is  also large. Then,  the corresponding  eigenvalues    
are bounded from below by ${\cal E} \rightarrow \infty$, i.e. 
$a \rightarrow -\infty$.

In this way we have obtained the complete spectrum for the zero modes 
${|\alpha,\, N, \, 0  \rangle}_B$.  
For a given $L$, we find numerically all the eigenvalues $a_{\alpha,\, N, \, 0 }$ of  Eq.(\ref{FDE}). 
For each of these eigenvalues, we obtain the energy by (\ref{AVE}) and  
the eigenfunctions $F_{\alpha,\, N, \, 0 }(c) $ by (\ref{SOLU}), together with the boundary conditions. The normalization constant can be fixed  
by the scalar product (\ref{scalarp}).

Among the zero modes,  we now focus on  the local minimum
($\alpha=0$) energy  states: 
${|0,\, N, \, 0  \rangle}_B$, for each  chiral sector $N$ of the theory. They have energies
${\cal E}_{0,\, N,\, 0}$.
An important consequence of the compactification prescription is that these states  
are not fully degenerated as they were in the non-compact model, which led to the introduction of the $\theta$-vacuum. Only  the pairwise degeneracy 
${\cal E}_{0,\, N,\, 0}={\cal E}_{0,\, -N,\, 0},\, N\neq 0$ remains. The only possibility involving full degeneracy 
arises in   
the limit $L \rightarrow \infty$. In this case ${\cal E}_{0,\, N,\,0}=0$ for all  
$N$.  

Most importantly, from  the numerical calculation we find that the absolute minimum value of 
${\cal E}_{0,\, N,\, 0}$ correspond to $N=0$. Thus, in the compact case the physical, 
non-degenerated, vacuum of the theory is $|0,\, 0, \, 0  \rangle_B$, so we  do not need to introduce  a  $\theta$-vacuum for the compact
Schwinger model.

\subsection{ The excited states}

In the previous section we have constructed the zero modes of the problem (no fermionic excitations),  
for each  chiral sector labeled by $N$. The corresponding   
excited states 
are obtained by applying  the creation 
operators $({ j}_\pm{}^m)^\dagger$ to them. Each 
individual action raises the energy by ${\cal E}_m$, as can be seen from Eq.(\ref{HTOTAL}). The excited states will be   
labeled by 
\begin{equation} 
\label{STATES} 
|\alpha,\ N, \ N_1,\ \dots, N_k, \dots \rangle_B, 
\end{equation} 
where $N_k$ is the total number of times that the operators 
$({ j}_\pm{}^k)^\dagger$ have 
been applied to the corresponding minimum energy 
state. This is the occupation number of the $k$-level. 
The zero modes correspond to $|\alpha,\ N, \ 0,\ \dots, 0, \dots \rangle_B$, i.e.  
$N_1= \dots = N_k \dots = 0$. These states were previously denoted by  
$ |\alpha,\ N, \ 0\rangle_B$.  
The total energy of the state (\ref{STATES}) is given by 
\begin{equation} 
{\cal E}_{\alpha,\, N, \,  N_1, \, N_2, \, \dots \, N_k,\,  \dots}= {\cal E}_{\alpha,\, N, \, 0}+\sum_{k>0} 
N_k \ {\cal E}_k. 
\end{equation} 
 
Since the values of ${\cal E}_{\alpha,\, N, \, 0}$ are not regularly spaced, we expect only  
accidental degeneracies, with the possible exception of the $L\rightarrow \infty$ limit, where the minimum energy states of the zero mode become degenerated with $a_{0,\, N,\, 0}=0$. 
In this case  we would  need to introduce the $\theta$- 
vacuum, in a similar form as in the non-compact case. Nevertheless, even in this situation 
we would not recover the standard non-compact case unless we further change the boundary conditions 
(\ref{BCWF}) and  (\ref{BCDWF}) to those employed in Refs. \cite{Manton,Link}.

Another interesting limit is $L \rightarrow 0$, where the standard harmonic oscillator spectra is recovered in the  
$N=0$ potential. Nevertheless, the   
$N \neq 0$  spectrum reduces to degenerate levels with  infinite energies. In other words, the non-compact case is neither recovered in  
this limit.   
 
In  this section we have shown that the  compact Schwinger model, which 
naturally arises from the loop 
approach to this problem, is also exactly soluble.  
 
\section{THE CHIRAL CHARGE} 
 
The fact that the chiral charge  $\bar Q_5$ is conserved in the full Hilbert space of the model is a direct  
consequence of the way in which  the Hilbert space has been constructed. 
We summarize the principal steps that  
lead  to this result: 
\begin{enumerate} 
\item{} First  we defined  $\bar Q_5$  in a gauge invariant way, in Eq.(\ref{MCC}).  
\item{} Using  the  operators $j_{+-}^{NN}$, $j_{-+}^{NN}$, we built the states (\ref{VED}) having minimum energy for each  
different label $2N$ of the chiral charge. 
\item{} The complete Hilbert space for the fermionic sector was 
constructed via the application of the raising operators $(j^n_+)^\dagger, (j^n_-)^\dagger $ upon the chiral states of item 2. Furthermore, the commutators (\ref{CRROFJ}) were calculated for the regularized currents, which validated the conservation of ${\bar Q}_5$ in such Hilbert space. 

\item{} The full Hilbert space of the model was  constructed in the Bogoliubov rotated frame, starting from the zero modes together with the same raising operators
$(j^n_+)^\dagger, (j^n_-)^\dagger $ employed in the fermionic sector.  This rotation did not affect ${\bar Q}_5$ thus preserving the commutation relations (\ref{ONCC}), (\ref{OSCC}).

\item{} Finally we have to establish the commutation relations among the electromagnetic
operators $c,\,  \frac{\partial}{\partial\, c}$ and the regularized currents $j_\pm{}^n$. These
currents depend upon $c$ only through the regularization factors ${\lambda}_{m,\, s}= |\epsilon_n|^s$. Thus, 
$c$ commutes with any of such currents. Up to this level, we see that  ${\bar Q}_5$ commutes with all the terms in the full Hamiltonian (\ref{HTOTAL}), except for the derivative
term which we analyze in the sequel. Let us consider the commutator

\be
\label{COMRP}
C_+{}^n= \left[\frac{\partial}{\partial\, c}, \, j_+{}^n\right]= 
{\rm lim}_{s\rightarrow 0} 
\sum_{m=-\infty}^{\infty} \frac{\partial}{\partial\, c}\left( \frac{1}{|\epsilon_n|^s}\right)a^\dagger_{m} a_{m+n}
\ee
First, let us consider the action of $C_+{}^n, \, n\neq 0$ upon an arbitrary vector
\be
|\{m_i \}\rangle=\prod_i \, a_{m_i}{}^\dagger\, |0\rangle
\ee
in the positive-chirality fermionic Fock  subspace. In general, the subindex $m_i$ will take values
over an infinite subset of integer numbers. The only  non-zero result of the action of  the $i^{th}$-term of (\ref{COMRP}) upon the above vector,  is to replace the $m_i +n$ fermion by the $m_i$
fermion, thus leading to a sum of linearly independent states. In this way, the $s\rightarrow 0$ limit must be taken  separately in  each term of the series and no infinite summation occurs. Since
\be
\frac{\partial}{\partial\, c}\left( \frac{1}{|\epsilon_n|^s}\right)\approx \, - \, \frac{s}{|\epsilon_n|^{s+1}},
\ee 
this limit is zero and the operators commute. 

Now, let us consider the $n=0$ case together with the action of $C_+{}^0$ upon the
local ground state $|N\rangle_B=F_N(c) \times \left| {\varepsilon}_N,2N \right\rangle$ of each chirality sector. We obtain
\ba
C_+{}^0\,|N\rangle_B&=& {\rm lim}_{s\rightarrow 0} 
\sum_{m=-\infty}^{N-1} \frac{\partial}{\partial\, c}\left( \frac{1}{|\epsilon_n|^s}\right)\, |N\rangle_B\nonumber \\
&=&
-\, \frac{e\, L}{2\, \pi}\, {\rm lim}_{s\rightarrow 0}\,  s\, \zeta(s+1, \, \frac{1}{2} +
\frac{e\,c\, L}{2\,\pi}-N)\, |N\rangle_B= -\, \frac{e\, L}{2\, \pi}\, |N\rangle_B, 
\ea 
where $\zeta(s,\, q)$ is the standard Riemann  zeta-function. We have used the property
${\rm lim}_{s\rightarrow 0}\,  s\, \zeta(s+1, \, q)=1$.
\cite{Riemann}

In analogous manner we consider
\be
\label{COMRM}
C_-{}^n
= \left[\frac{\partial}{\partial\, c}, \, j_-{}^n\right]= 
{\rm lim}_{s\rightarrow 0} 
\sum_{m=-\infty}^{\infty} \frac{\partial}{\partial\, c}\left( \frac{1}{|\epsilon_n|^s}\right)b^\dagger_{m+n} b_{m}.
\ee
Again, the action of $C_-{}^n, \, n\neq 0$ upon an arbitrary state
$|\{n_i \}\rangle=\prod_i \, b_{n_i}{}^\dagger\, |0\rangle
$ is zero. For $n=0$ we obtain, upon the local ground states,  
\ba
C_-{}^0\,|N\rangle_B&=& {\rm lim}_{s\rightarrow 0} 
\sum_{m=N}^{\infty} \frac{\partial}{\partial\, c}\left( \frac{1}{|\epsilon_n|^s}\right)\, |N\rangle_B \nonumber \\
&=&
\, \frac{e\, L}{2\, \pi}\, {\rm lim}_{s\rightarrow 0}\,  s\, \zeta(s+1, \, \frac{1}{2} -
\frac{e\,c\, L}{2\,\pi}+ N)\, |N\rangle_B = \, \frac{e\, L}{2\, \pi}\, |N\rangle_B, 
\ea

The above results lead to
\be
\left[\frac{\partial}{\partial\, c}, \, {\bar Q}_5  \right]\, |N\rangle_B= \left( C_+{}^0\,  - C_-{}^0\,  + \frac{e\, L}{\pi}\right) |N\rangle_B = 0. 
\ee
Besides, any excited state is constructed by applying the raising operators $(j_\pm{}^n)^\dagger,  n\geq1$ to $|N\rangle_B$. These operators commute with ${\bar Q}_5$ and $\frac{\partial}{\partial\, c}$ in such way that the commutator $\left[\frac{\partial}{\partial\, c}, \, {\bar Q}_5  \right]$ is zero in the full Hilbert space of the problem. This completes our proof that ${\bar Q}_5 $ commutes with the total
Hamiltonian (\ref{HTOTAL}).

\end{enumerate}

Since $\bar Q_5$ and $Q$ are conserved and gauge invariant in the compact model, we would 
expect to obtain zero values for the fermionic condensates  
$\langle \bar \psi \psi \rangle_{0,\,0,\,0}$ and  
$\langle \bar \psi \gamma_5 \psi \rangle_{0,\,0, \,0}$ which measure the 
corresponding amounts of symmetry breaking. Indeed this result is obtained  here, 
as a consequence of the physical vacuum been non-degenerated.  We perform the calculation in the original unrotated frame, where the wave function is $|0, \, 0, \, 0\rangle= U
|0, \, 0, \, 0\rangle_B$. In this frame, the fermionic 
bilinears are  
\begin{eqnarray} 
 \bar \psi \psi &=& 
\frac{1}{L} \sum_{n \in Z} \sum_{m \in Z} e^{-i\frac{2\pi n}{L}x}  
\left( j_{-+}^{m+n,m} + j_{+-}^{m,m-n}\right)  \nonumber \\ 
\bar \psi \gamma_5 \psi &=& 
\frac{1}{L} \sum_{n \in Z} \sum_{m \in Z} e^{-i\frac{2\pi n}{L}x}\left(  
j_{-+}^{m+n,m} - j_{+-}^{m,m-n}\right). 
\end{eqnarray} 
 
For example, in the case of the bilinear $\bar \psi \psi$, the calculation 
goes as follows 
\begin{eqnarray} 
 \langle 0,\,0,\,0\,|\bar \psi \psi |\, 0,\,0,\,0 \,\rangle &=&
{}_B\langle 0,\,0,\,0\,|\, U^\dagger\, \bar \psi \psi \, U |\, 0,\,0,\,0 \,\rangle_B =  
 \langle \varepsilon_0,0 |\, U^\dagger \, \bar \psi \psi\, U \, |\varepsilon_0,0 \rangle\nonumber \\ 
&=&\frac{1}{L} \sum_{n \in Z} \sum_{m \in Z} e^{-i\frac{2\pi n}{L}x} 
\langle \varepsilon_0,0 |\, U^\dagger \, \left( j_{-+}^{m+n,m} + j_{+-}^{m,m-n}\right)\,U \, |\varepsilon_0,0 \rangle\nonumber \\ 
&=&\frac{1}{L} \sum_{n \in Z} \sum_{m \in Z} e^{-i\frac{2\pi n}{L}x}\, \frac{1}{2}\, 
\langle \varepsilon_0,0 |\, U^\dagger \, \left(- \left[{\bar Q}_5,\, j_{-+}^{m+n,m}\right] + \left[{\bar Q}_5,\, j_{+-}^{m,m-n}\right]\right)\,U \, |\varepsilon_0,0 \rangle\nonumber \\ 
&=& 0, 
\end{eqnarray}
where we have assumed unit normalization for the  electromagnetic part of the wave
function. We have also used the appropriate commutators in (\ref{OSCC}) to rewrite $j_{-+}^{m+n,m}$ and $ j_{+-}^{m,m-n} $. The null result follows since the expectation value of each  commutator in the above summation is zero. This is  because ${\bar Q}_5$ 
commutes with  $U, \, U^\dagger$ and the state $|\varepsilon_0,0 \rangle$ has definite chiral charge (zero in our conventions). A similar calculation can be performed for the chiral condensate. 


\section{Summary and Conclusions } 
 
Motivated by the loop-space formulation of $1+1$ QED, we have exactly solved the
compact Schwinger model, defined by the condition that both the spatial coordinate $x$ together with
the electromagnetic degree of freedom $c$ behave as angular variables. In other words, we are dealing with compact $U(1)$ as the corresponding gauge group.  Many approaches to
the standard (non-compact) Schwinger model also start from a compacted electromagnetic variable, but the  boundary conditions are chosen  in such a way that this degree of freedom is ultimately extended to  the line \cite{Manton,HetHo,Shifman,Link,IsoMurayama}.

We maintain the angular character of the electromagnetic variable. Consistency with this
requirement  leads to major differences between this model and the standard one. The most remarkable are: different transformations properties under gauge transformations,  different spectra and wave functions and the existence of  a conserved gauge invariant chiral charge. 
Nevertheless, the basic features which allow for the exact solution remain the same: the Sugawara transformation of the originally
linear fermionic piece of the Hamiltonian and the Bogoliubov rotation of the full Hamiltonian.
The non-conventional results we have obtained  can be hardly surprising if we think of the situation as a more sophisticated analogy of a given differential equation subjected to different boundary conditions, dictated by different choices of the topology in the corresponding space.

The first consequence of the compactification  is that the surviving electromagnetic degree of freedom $c$ is invariant under both SGT and LGT. All further properties of the compact model follow  directly from this invariance. In particular, it implies the gauge invariance of the individual eigenvalues
$\epsilon_n$ of the Hamiltonian in the fermionic Fock space, which subsequently leads to the full gauge invariance  the  fermionic
operators $a_n, \, b_n $. These properties have to be contrasted with the non-compact case, where
$\epsilon_n \rightarrow \epsilon_{n+1}$ and  $\, a_n \rightarrow a_{n+1}, \, b_n \rightarrow
b_{n+1}$ under LGT.

The next important consequence has to do with the definition of the total chiral charge. In both cases one starts with  $Q_5$ defined in Eq. (\ref{DEFCH}), which is not conserved thus leading to the standard axial-charge anomaly. This chiral charge is  invariant(non-invariant) under LGT in the compact(non-compact case). Next one introduces the modified chiral charge ${\bar Q}_5$, defined in Eq. (\ref{MCC}), which is conserved and independent of the electromagnetic degree of freedom
$c$ in both cases. The modified chiral charge ${\bar Q}_5$  retains the invariance(non-invariance) under LGT in the compact(non-compact) case, this time in virtue of the transformation properties of the fermionic operators. Thus, the compactification requirement allows us to have  the conservation of the 
electric charge together with  the modified chiral charge, leading to the absence of both the vector and axial-vector charge anomalies. Consequently the  condensates $\langle {\bar \psi}\, \psi \rangle$ and
$\langle {\bar \psi}\,\gamma_5\,  \psi \rangle$ are zero in the compact case. 

Nevertheless, the axial-current anomaly is also present in the compact case, as we now
discuss. The charge 
$Q_5$ arises from the current
\be 
J_{5\, \mu}(x)={\bar \psi}(x)\gamma_\mu\, \gamma_5 \, \psi(x)=\frac{1}{L}\sum_{n=-\infty}^{n= +\infty}\, {\rm e}^{- \frac{2\,\pi\,i\, n}{L}x} J_{5\,\mu}{}^n, 
\ee
where
\ba
\label{ACX}
J_{5\,0}{}^n= j_+{}^n - \left(j_-{}^n\right)^\dagger, \qquad  J_{5\,1}{}^n= j_+{}^n + \left(j_-{}^n\right)^\dagger.
 \ea
The operators $j_+{}^n $ and $j_-{}^n $, in their regulated forms, are non-invariant under LGT in the non-compact case. They are fully gauge invariant in the compact case. In both cases
this current possesses the anomaly
\be
\label{SJ5AN}
\partial^\mu\, J_{5\, \mu}=-\, \frac{e}{\pi}\, E(x), 
\ee
which can be directly calculated using the expression (\ref{ACX}) together with the unrotated Hamiltonian
(\ref{HM}) and the Gauss law (\ref{AFORE}).

On the other hand, one can  introduce the conserved  local  current 
\be
\label{J5BAR}
{\bar J}_{5\, \mu}(x)=J_{5\, \mu}(x)-\frac{e}{\pi}\, \epsilon_{\mu\nu}\, A^\nu, \qquad
\epsilon_{01}=+1.
\ee
leading to the charge
\be
{\bar Q}_5=\int_0^L dx \,{\bar J}_{5\, 0}= Q_5 + \frac{e\, c\, L}{\pi}.
\ee
 Nevertheless, the current (\ref{J5BAR}) is not gauge invariant, either in the non-compact or in the compact cases.  In this way, it cannot be restricted to the physical Hilbert space of the problem. 

Summarizing this point, the axial current anomaly (\ref {SJ5AN}) is also present in the compact
Schwinger model, and it cannot be removed, in spite it is possible to define the conserved and gauge invariant modified chiral charge 
${\bar Q}_5$.

Next we discuss the spectra of the models. In the standard non-compact case we have an infinite
set of sectors labeled by the integer $N$, which are connected by LGT. The corresponding zero
modes in each sector have energies given by $(n + 1/2)\, e/\sqrt{\pi}\,,  n=0,1,2, \dots$ , independently  of the label $N$, thus been infinitely degenerated. It is precisely this property that requires  the introduction of the $\theta$-vacuum. 

In the compact case, the sectors labeled by $N$, corresponding here to the eigenvalues $2\, N $ of ${\bar Q}_5$, are also present. They are connected through the operators $j_{+-},\, j_{-+} $.  Nevertheless, due to the boundary conditions (\ref{BCWF}) and (\ref{BCDWF})  the corresponding zero modes
energies depend upon the label $N$ and are non-degenerated as can be seen in Figs. 2 and 3.
In fact, the lowest energy state corresponding to the $N=0$ sector is the ground state of the
model. Thus, no $\theta$-vacuum is required in the present case. 

The exited states are constructed by the same procedure in both cases: by applying the raising operators $(j_{\pm}{}^m)^\dagger$ to the zero mode states. Nevertheless, their action produce both
eigenvectors and eigenvalues which are different with respect to the non-compact case. The non-equally spaced spectra of the zero mode does not lead to a particle interpretation of the compact model, as it is the case in  the non-compact situation.  

Finally we comment once again that, for a given $L\neq0$, the boundary conditions  for the compact model (Eqs.(\ref{BCWF}) and (\ref{BCDWF}))
and those of the non-compact case ( Eqs. (3.15) of Ref. \cite{Manton}, or Eq. (48) of Ref.
\cite{Link}) can not be continuously connected between each other. Thus, neither model can 
be
obtained from the other through an adequate limiting process, emphasizing once again that the
compactification condition has produced  a Schwinger model which is different from the standard
one.

\vskip.7cm

\centerline{\bf THE APPENDIX} 

\vskip.4cm
 
 To begin with we show how  the commutation relations 
(\ref{CRROFJ}), valid  on the fermionic Fock space, are obtained. We prove this for the 
currents in the positive chiral sector. The negative chiral sector  
case is analogous. The commutator among the regularized currents is 
\begin{equation} 
\label{BASCOM} 
[ j_+{}^n,   (j_+{}^m)^\dagger ]= 
                                       \lim_{s\to 0 \atop r\to 0} 
      \sum_p \left ( \frac{1}{\lambda_{p+m,s}\lambda_{p+n,r}}- 
      \frac{1}{\lambda_{p,s+r}} \right ) a^\dagger_{p+m} a_{p+n}, 
\end{equation} 
Acting the 
commutator 
on the vacuum leads to 
\begin{equation} 
\label{COMVAC} 
[ j_+{}^n,   (j_+{}^m)^\dagger ]\left| {\cal E}_N, 2N \right \rangle = 
\lim_{s \to 0 \atop r\to 0} 
 \sum_{p=-\infty}^{N-1-n} \left ( 
\frac{1}{\lambda_{p+m,s}\lambda_{p+n,r}}- 
              \frac{1}{\lambda_{p,s+r}} \right ) 
  a^\dagger_{p+m} a_{p+n} \prod_{t=-\infty}^{N-1}a_t^\dagger |0 \rangle 
\otimes \prod_{t=N}^{\infty}b_t^\dagger |0 \rangle. 
\end{equation} 
Without loss 
of generality,  in the following we take $n$ as a positive integer and we separate the calculation into three parts: 
 
(i) $n=m$: here we obtain 
\begin{equation} 
\label{MINVAC} 
  [ j_+{}^n,   (j_+{}^m)^\dagger ]\left |{\cal E}_N,2N \right \rangle = 
   \sum_{p=N-n}^{N-1} \left |{\cal E}_N,2N \right \rangle   
   = n \left |{\cal E}_N,2N \right \rangle . 
\end{equation} 
 
(ii) $n>m$: since here $p+m < p+n$ for every $p$, we have that each term in the summation (\ref{COMVAC}) will always contain a repeated fermionic creation operator. Then we obtain 
\begin{equation} 
\label{MMENVAC} 
[ j_+{}^n,   (j_+{}^m)^\dagger ]\left |{\cal E}_N,2N \right \rangle= 0 . 
\end{equation} 
 
(iii) $n<m$: here we need to be more careful because we are left with a 
finite sum that vanishes after taking the corresponding limits 
\begin{eqnarray} 
\label{MMANVAC} 
[ j_+{}^n,   (j_+{}^m)^\dagger ]\left |{\cal E}_N,2N \right \rangle=&\lim_{s\to 0 \atop r\to 0} 
 \sum_{p=N-m}^{N-1-n} \left ( \frac{1}{\lambda_{p+m,s}\lambda_{p+n,r} }- 
              \frac{1}{\lambda_{p,s+r}} \right )\nonumber \\ 
& \times a^\dagger_{p+n} a_{p+n} \prod_{t=-\infty}^{N-1}a_t^\dagger |0 
\rangle \otimes \prod_{n=N}^{\infty}b_n^\dagger |0 \rangle= 0. 
\end{eqnarray} 
 
However, this is not the end of the story because we want to make sure 
that 
the commutation relations (\ref{CRROFJ}) are valid, not only  when acting on the vacuum, but in the whole Fock 
space. In other words, we need show that 
\begin{equation} 
\label{COMWFFS} 
[ j_+{}^n,   (j_+{}^m)^\dagger ] 
\prod_{p=1}^{k} (j_+{}^{i_p})^\dagger 
\left |{\cal E}_N,2N \right \rangle = 
 n \delta_{n,m} \prod_{p=1}^{k} (j_+{}^{i_p})^\dagger 
\left |{\cal E}_N,2N \right \rangle,                                                       \end{equation} 
for an arbitrary number $k$ of currents acting on the vacuum. This can be done by induction. 
For $k=1$ we have 
\begin{equation} 
[ j_+{}^n,   (j_+{}^m)^\dagger ] \left ( (j_+{}^{i_1})^\dagger 
   \left |{\cal E}_N,2N \right \rangle \right ) = 
 [[j_+{}^n,   (j_+{}^m)^\dagger ], (j_+{}^{i_1})^\dagger ]  
\left |{\cal E}_N,2N \right \rangle 
 + n\delta_{n,m} \left ( (j_+{}^{i_1})^\dagger 
   \left |{\cal E}_N,2N \right \rangle \right). 
\end{equation} 
Now, it is easily shown that for the three different cases, 
$n>p+m,\quad n<p+m, \quad n=p+m,$ the first term in the RHS of the above 
equation is zero, thus proving the assertion (\ref{COMWFFS}) for $k=1$. 
 
Next we assume that (\ref{COMWFFS})is valid for $k$ currents and we prove that it is also true for $k+1$ of them. To this end, let us consider 
\begin{eqnarray} 
\label{GCOMVAC} 
 && \Bigl [ j_+{}^n,   (j_+{}^m)^\dagger \Bigr] 
 \prod_{p=1}^{k+1} (j_+{}^{i_p})^\dagger \left |{\cal E}_N,2N \right \rangle 
\nonumber \\ 
&& = \Biggl [ \cdots \biggl [ \Bigl [ [j_+{}^n,   (j_+{}^m)^\dagger ], 
(j_+{}^{i_{1}})^\dagger \Bigr ], 
(j_+{}^{i_{2}})^\dagger \biggr ],\cdots ,(j_+{}^{i_{k+1}})^\dagger 
\Biggr] 
\left |{\cal E}_N,2N \right \rangle \nonumber \\ 
 & &  \qquad + n \delta_{n,m} \prod_{p=1}^{k+1} (j_+{}^{i_p})^\dagger 
    \left |{\cal E}_N,2N \right \rangle. 
\end{eqnarray} 
The first term in the RHS of the above equation is 
\begin{eqnarray} 
  \Biggl [ \cdots && \biggl [ \Bigl [ [j_+{}^n,   (j_+{}^m)^\dagger ], 
(j_+{}^{i_{1}})^\dagger \Bigr ], 
(j_+{}^{i_{2}})^\dagger \biggr ],\cdots ,(j_+{}^{i_{k+1}})^\dagger 
\Biggr] 
 \nonumber \\ 
&& =\lim_{r,s \to 0 \atop t_1, \cdots ,t_k \to 0} 
\sum_q C_k(q,n,m,i_1, \cdots ,i_k) 
a^\dagger_{q+m+ i_1+ \cdots + i_k} a_{q+n}, 
\end{eqnarray} 
where the quantities $C_k(q,n,m,i_1, \cdots ,i_k)$  
are defined in terms of their recursion relations by  
\begin{eqnarray} 
 C_k(q,n,m,i_1, \cdots ,i_k)&=& 
C_{k-1}(q+i_k,n,m,i_1, \cdots ,i_{k-1})\frac{1}{\lambda_{q+n,t_k}}\nonumber \\ 
   & & \qquad  - 
C_{k-1}(q,n,m,i_1, \cdots ,i_{k-1}) 
 \frac{1}{\lambda_{q+m+i_1+ \cdots +i_{k-1},t_k}}, \nonumber \\ 
C_0(q,n,m)&=&\frac{1}{\lambda_{q+m,s} \lambda_{q+n,r}} - 
           \frac{1}{\lambda_{q,s+r}}. 
\end{eqnarray} 
From the above, it is possible to show that in  the three cases 
$n (>,<,=) m+i_1 +\cdots+ i_k $, we have  
\begin{equation} 
\Biggl [ \cdots \biggl [ \Bigl [ [j_+{}^n,   (j_+{}^m)^\dagger ], 
(j_+{}^{i_{1}})^\dagger \Bigr ], 
(j_+{}^{i_{2}})^\dagger \biggr ],\cdots (j_+{}^{i_{k+1}})^\dagger 
\Biggr] 
\left |{\cal E}_N,2N \right \rangle =0  , 
\end{equation} 
Thus the proof is complete. 

Next we verify that the regulated currents (\ref{REGJ}) satisfy the hermiticity
property $(j_{\pm}{}^n)^\dagger=j_{\pm}{}^{-n}$. Let us concentrate in a $+$ current and  define the operator
\be
D_+{}^n=(j_{+}{}^n)^\dagger- j_{+}{}^{-n}={\rm lim}_{s\rightarrow 0} 
\sum_{m=-\infty}^{\infty} \left( \frac{1}{\lambda_{m-n,\, s}}- \frac{1}{\lambda_{m ,\, s}}\right)a^\dagger_{m} a_{m-n}.
\label{DIFO}
\ee
Next we apply the above operator to an arbitrary vector
\be
|\{m_i \}\rangle=\prod_i \, a_{m_i}{}^\dagger\, |0\rangle
\ee
in the positive-chirality fermionic Fock  subspace. In general, the subindex $m_i$ will take values
over an infinite subset of integer numbers.

For $n=0$ the regulated operator (\ref{DIFO}) is trivially zero, so that we concentrate in the
$n \neq 0$ case. Here, only the values $m \in \{m_i + n \}$ give a non-zero result. The action in each term $m$ of the sum (\ref{DIFO}) is to replace the $m_i$ fermion in the state by an $m_i + n$ fermion. In this way,  the resulting vectors are linearly independent and the
$s\rightarrow0$ limit has to be taken separately  in any of these contributions, leading to zero in each case. In other words, the only infinite sum that could have appeared corresponds to the $n=0$ case. The proof for the negative-chirality sector follows along  the same lines.

\acknowledgments 
Partial support from the grants CONACyT  32431-E and DGAPA-UNAM-IN100397 is 
acknowledged.

\begin{figure}[h] 
\centerline{\psfig{figure=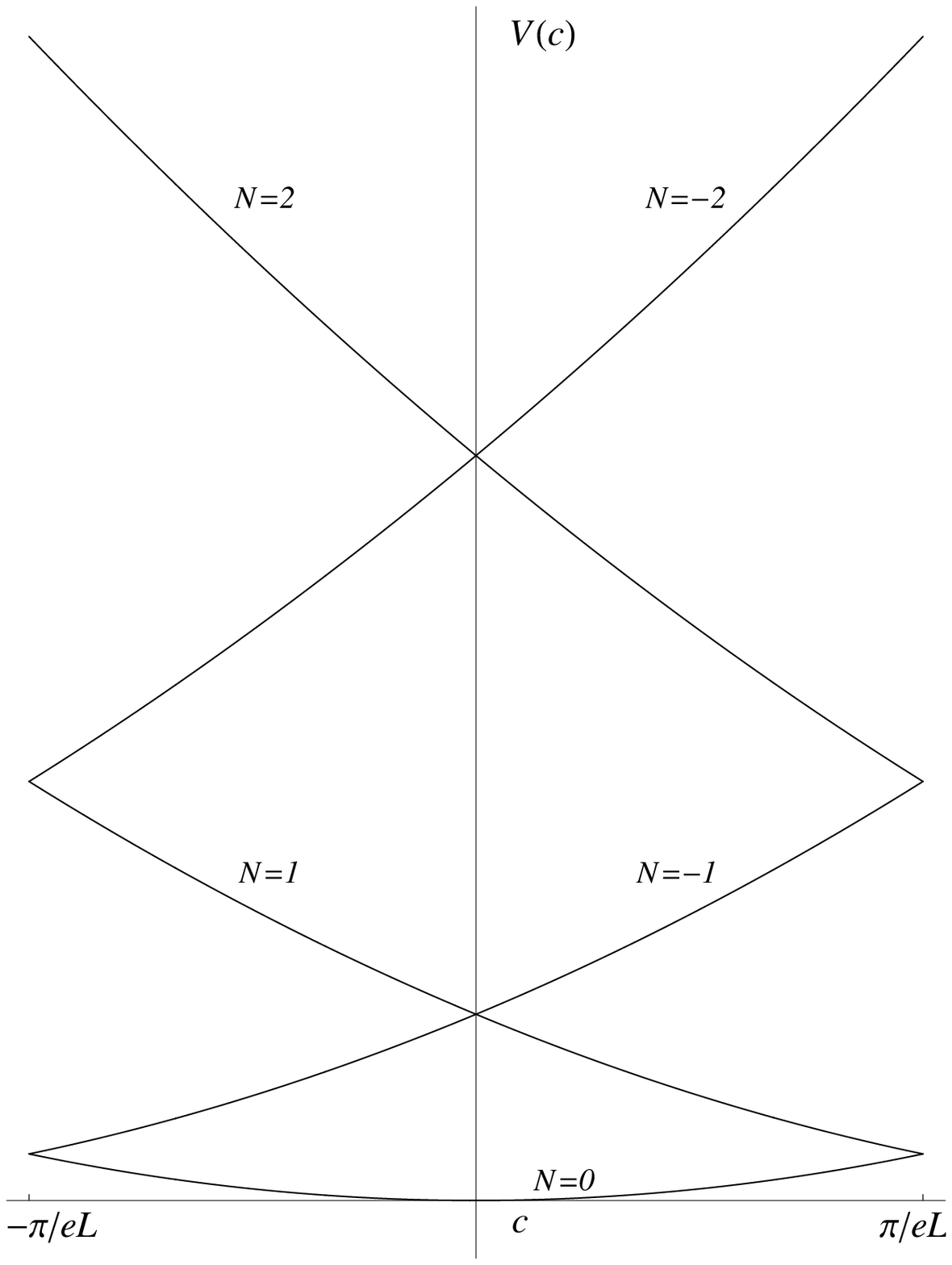,width=8in,height=8in}} 
\caption{The potentials $V_N(c)$ in the compact Schwinger model} 
\label{VEG1} 
\end{figure}

\begin{figure}[h] 
\centerline{\psfig{figure=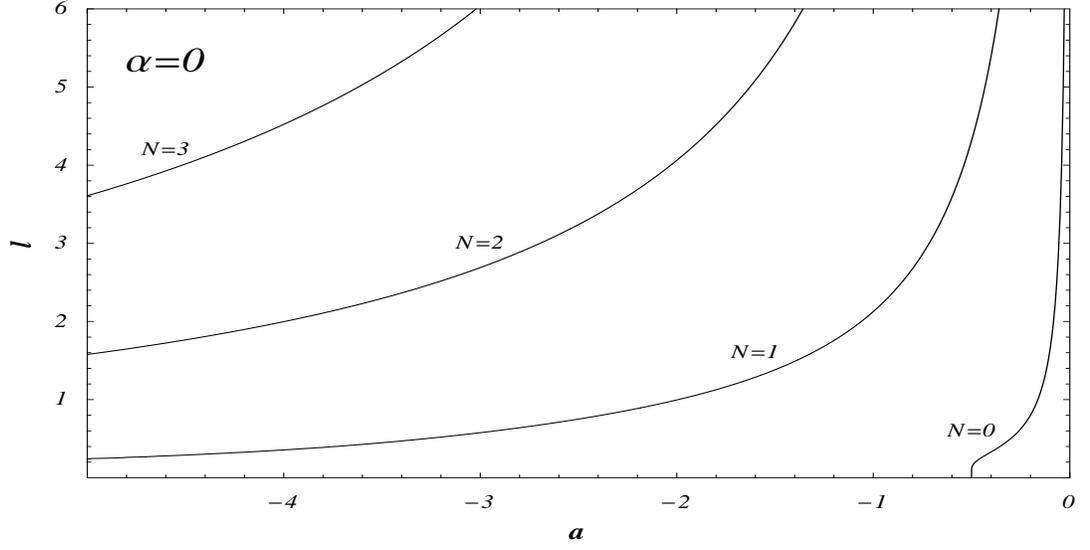,width=8in,height=8in}} 
\caption{ The 
numerical solution for the parameter $a_{0,N}(l)$,  for $N=0, 1, 2, 3$ and 
a given value of $l$. The energies are 
${\cal E}_{0,N,0}=-(e/\pi^{1/2})\, a_{0,N}$}
\label{VEG2} 
\end{figure}

\begin{figure}[h] 
\centerline{\psfig{figure=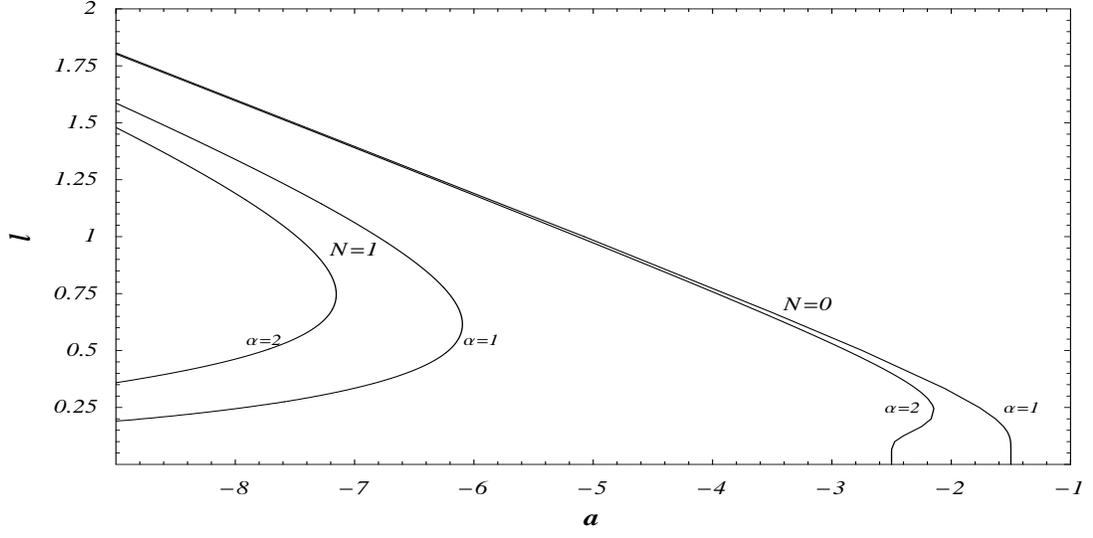,width=8in,height=8in}} 
\caption{ The 
numerical solution of $a_{\alpha,N}(l)$, with $N=0,1$ and
$\alpha=1,2$, for a given value of $l$, is given.
The energies are 
${\cal E}_{\alpha,N,0}=-(e/\pi{1/2})\, a_{\alpha,N}$}
\label{VEG3} 
\end{figure}

\end{document}